\documentclass{emulateapj}

\def\arcdeg{\hbox{$^\circ$}}

\def\arcsec{\hbox{$^{\prime\prime}$}}
\def\deg2{\hbox{$\rm deg^{2}$}}

\def\lsim{\mathrel{\rlap{\lower4pt\hbox{\hskip1pt$\sim$}}\raise1pt\hbox{$<$}}}                
\def\gsim{\mathrel{\rlap{\lower4pt\hbox{\hskip1pt$\sim$}}\raise1pt\hbox{$>$}}}                

\lefthead{Drake et~al.}
\righthead{Ultra-short Period Binaries} 

\begin{document}
\title{Ultra-short Period Binaries from the Catalina Surveys}

\author{
A.J.~Drake\altaffilmark{1}, S.G.~Djorgovski\altaffilmark{1}, D. Garc\'{\i}a-\'Alvarez\altaffilmark{3,4,5}, 
M.J.~Graham\altaffilmark{1}, M.~Catelan\altaffilmark{2},\\ A.A.~Mahabal\altaffilmark{1},
C.~Donalek\altaffilmark{1}, J.L.~Prieto\altaffilmark{6}, G.~Torrealba\altaffilmark{2}, 
S.~Abraham\altaffilmark{7},\\ R.~Williams\altaffilmark{1}, S.~Larson\altaffilmark{8}, 
and E.~Christensen\altaffilmark{8}
}

\altaffiltext{1}{California Institute of Technology, 1200 E. California Blvd, CA 91225, USA}
\altaffiltext{2}{Pontificia Universidad Cat\'olica de Chile, Departamento de Astronom\'ia y Astrof\'isica, 
Facultad de F\'{i}sica, Av. Vicu\~na Mackena 4860, 782-0436 Macul, Santiago, Chile}
\altaffiltext{3}{Instituto de Astrof\'{\i}sica de Canarias, Avenida V\'{\i}a L\'actea,
38205 La Laguna, Tenerife, Spain}
\altaffiltext{4}{Departamento de Astrof\'{\i}sica, Universidad de La Laguna,
38205 La Laguna, Tenerife, Spain}
\altaffiltext{5}{Grantecan S.\,A., Centro de Astrof\'{\i}sica de La Palma, Cuesta de San Jos\'e,
38712 Bre\~na Baja, La Palma, Spain}
\altaffiltext{6}{Department of Astronomy, Princeton University, 4 Ivy Ln, Princeton, NJ 08544}
\altaffiltext{7}{St. Thomas College, Kozhencheri - 689641, India}
\altaffiltext{8}{The University of Arizona, Department of Planetary Sciences,  Lunar and Planetary Laboratory, 
1629 E. University Blvd, Tucson AZ 85721, USA}

\begin{abstract} 
  
  We investigate the properties of 367 ultra-short period binary candidates selected from 31,000 sources
  recently identified from Catalina Surveys data. Based on light curve morphology, along with WISE, SDSS and GALEX
  multi-colour photometry, we identify two distinct groups of binaries with periods below the 0.22 day contact binary
  minimum.  In contrast to most recent work, we spectroscopically confirm the existence of M-dwarf+M-dwarf contact
  binary systems.  By measuring the radial velocity variations for five of the shortest-period systems, we find examples
  of rare cool-white dwarf+M-dwarf binaries. Only a few such systems are currently known. Unlike warmer
  white dwarf systems, their UV flux and their optical colours and spectra are dominated by the M-dwarf companion.  We
  contrast our discoveries with previous photometrically-selected ultra-short period contact binary candidates, and
  highlight the ongoing need for confirmation using spectra and associated radial velocity measurements.  Overall, our
  analysis increases the number of ultra-short period contact binary candidates by more than an order of magnitude.

\end{abstract}

\section{Introduction}

The study of eclipsing binaries provides the opportunity of determining stellar parameters with a high degree of
accuracy using constraints on the geometry and motions of the system (Southworth 2012 and refs therein). Amongst 
other things, eclipsing binaries can be used to probe various stages of stellar evolution via the determination 
of system parameters such as the orbital period, inclination, radii and masses.

Types of eclipsing binaries are generally defined by the degree of separation of the components. These include
over-contact, contact, semi-detached and detached systems that can often be discerned from light curve shapes.
Eclipsing contact binaries are referred to as W Ursae Majoris (W UMa's) stars (or EW variable types). Such systems allow
mass to flow from one star to the other, and both stars usually have similar temperatures and types. Here the depth of
the eclipses is most dependent on the system geometry, while in contrast, the depth of the eclipses reflects the relative effect
temperatures in detached binaries (Rucinski 2001). Nevertheless, slight differences in the eclipse depth still occur in
contact systems due to differences in the temperature of the component stars. 

For many years it has been suspected that there is a minimum period for contact binaries at $P \sim\rm 0.22$ days
(Rucinski et al.~1992, 1997).  Numerous explanations have been put forth to explain the observed abrupt cut off. For
example, Stepien (2006) proposed that the limit was due to the decrease in the efficiency of angular momentum loss with
decreasing mass on the main sequence. In such cases the current age of the systems comprising the binary population
produces the limit. More recently, Jiang et al.~(2012) investigated mass transfer in low-mass binaries, and found this
to be unstable. They suggest that systems with the lowest mass might quickly coalesce, leading to their observational
absence.  

In the past few years, increasing numbers of eclipsing binary candidates have been discovered with periods 
less than 0.22 days.  For example, the detached binary, OGLE-BW03-V038 was discovered with a period of 0.1984 days by 
Maceroni \& Rucinski (1997), and a similar detached eclipsing binary system with main-sequence stars was found with a 
period of 0.1926 days by Norton et al.~(2007) and confirmed by Dimitrov \& Kjurkchieva (2010). However, as detached 
systems are not in the process of mass transfer, they exist in different evolutionary state than contact systems. 

Recently, Norton et al.~(2011) identified 14 new eclipsing systems with periods $P < \rm 0.22$ days among 30 million 
SuperWASP sources. Their results include a number of possible contact systems with periods in the 0.2 to 0.22 day range.
Similarly, Nefs et al.~(2012) discovered 14 eclipsing binary candidates with periods less than 0.22 days.
Among these systems Nefs et al.~(2012) spectroscopically confirm a detached system with a 0.18 day period 
containing an M-dwarf. However, they did not attempt to measure radial velocities.
In contrast, Davenport et al.~(2013) discovered a likely contact binary system with a period of 0.19856 days. 
They confirm the presence of a M-dwarf and find radial velocity variations that are consistent binary M-dwarf system.
Nevertheless, despite the increasing evidence for short period contact binary systems below the 0.22 day period cutoff,
only a handful of candidates have so far been spectroscopically confirmed, and fewer still have had their
component masses determined via radial velocities measurements. 

In this analysis we will investigate both the spectral types and radial velocities for ultra-short period binary systems
discovered during the compilation of the Catalina Surveys periodic variable stars catalog (Drake et al.~2014).

\section{Observational data}

The Catalina Sky Survey\footnote{http://www.lpl.arizona.edu/css/} uses three telescopes to cover the sky between
declination $\delta = -75$ and +65 degrees in order to discover Near-Earth Objects (NEOs) and Potential Hazardous
Asteroids (PHAs). Catalina observations do not cover regions within 15 degrees of the Galactic plane because
of blending in crowded stellar regions.  All the images are taken unfiltered, and photometry is carried out using the
aperture photometry program SExtractor (Bertin \& Arnouts 1996). These measurements are transformed to an approximate V
magnitude ($V_{CSS}$) using fiducial 2MASS colour-selected G-dwarf stars as outlined in Drake et al.~(2013).

In this paper, we select and analyze the ultra-short period systems ($P < 0.22$ days) among the 31,000 contact binary
candidates discovered by Drake et al.~(2014) from data in the Catalina Data Release-1 (CSDR1; Drake et al.~2012).  The
CSDR1 dataset itself consists only of data taken with the Catalina Schmidt Survey telescope.  This includes an average
of 250 observations per position spanning six years for 198 million discrete sources. The CSDR1 sources have $12 < V <
20$, and lie in the region $0\arcdeg < \alpha < 360\arcdeg$ and $-30\arcdeg < \delta < 65\arcdeg$. The lightcurve data
for all CSDR1 sources (including those presented here) is publicly available online\footnote{http://catalinadata.org}.

\section{Short Period Binaries}

The Drake et al.~(2014) variable catalog contains 367 systems with periods less than 0.22 days, and of these 73 have
periods less than 0.2 days.  The eclipsing systems were selected based on the morphological classification from among
61,000 periodic variables in the catalog. Additional confirmation of the variable classifications in the catalog was
performed by Drake et al.~(2014) using colour information derived by combining CSS, WISE and SDSS photometry.
Spectroscopic parameters were also derived from SDSS data and confirm the separation between the eclipsing binaries and
pulsating stars such as RR Lyrae stars.  As the photometry spans many years we do not expect contamination by spotted 
variable stars (such as RSCVns), since such objects exhibit clear changes in the average brightness, amplitude and
phase of flux variations (eg. see Drake 2006, figure 2 and Drake et al.~2014, figure 18).

\begin{figure}[ht]{
\epsscale{1.0}
\plotone{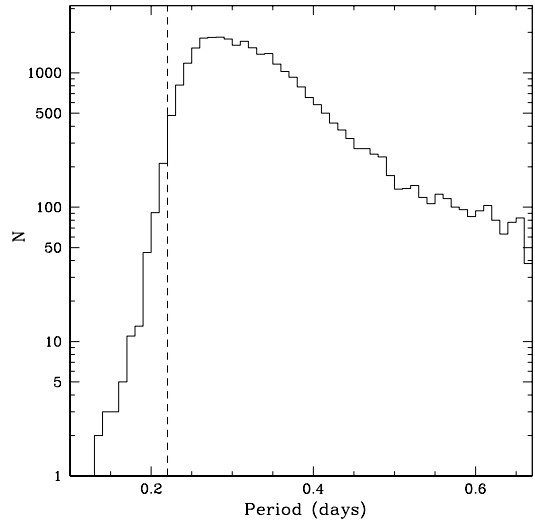}
\caption{\label{Bdist}
The period distribution of contact binaries.
The solid line give the distribution of contact binary
candidates from CSDR1. The dashed line denotes the location 
of the purported 0.22 day contact binary period minimum.
}
}
\end{figure}

In Figure \ref{Bdist}, we present the period distribution of the 31,000 Drake et al.~(2014) contact binary candidates.
Here we see there is indeed a very sharp decline in the number of short period systems below $\sim 0.27$ days. Given the
large number of contact systems in CSDR1, the fraction of ultra-short period objects amounts to only $\sim 0.26 \%$ of
the total number. This result in itself very strongly supports the scarcity of short period systems.  However, we note
that past surveys for such systems have had a clear observational bias against finding such systems.  That is to say,
the stars that comprise short-period systems are much fainter than the dG and dK stars that make up longer-period
systems.  For example, in $V$-band, an M0V star is 4.4 magnitudes fainter than a G0V star. Thus, shallow wide-field 
synoptic surveys, such as ASAS (Pojmanski 1997) that only reached V=13, have probed a very small volume of potential 
M-dwarf binaries compared to CSDR1 (e.g., see Rucinski 2006, figure 4).

In order to investigate the nature of the ultra-short period sources we reexamined the light curves of all the objects
Drake et al.~(2014) identified as eclipsing contact and ellipsoidal binary candidates below the nominal 
ultra-short period limit (0.22 day). We observed a clear division in the morphology of the source light curves.
Approximately half exhibited the W-shaped light curves expected for W UMa contact binaries, while the other half
exhibited much more sinusoidal light curves, as seen in over-contact systems where the stars share an outer envelope 
and both exceed their Roche limit.

\begin{figure*}[ht]{
\epsscale{1.0}
\plottwo{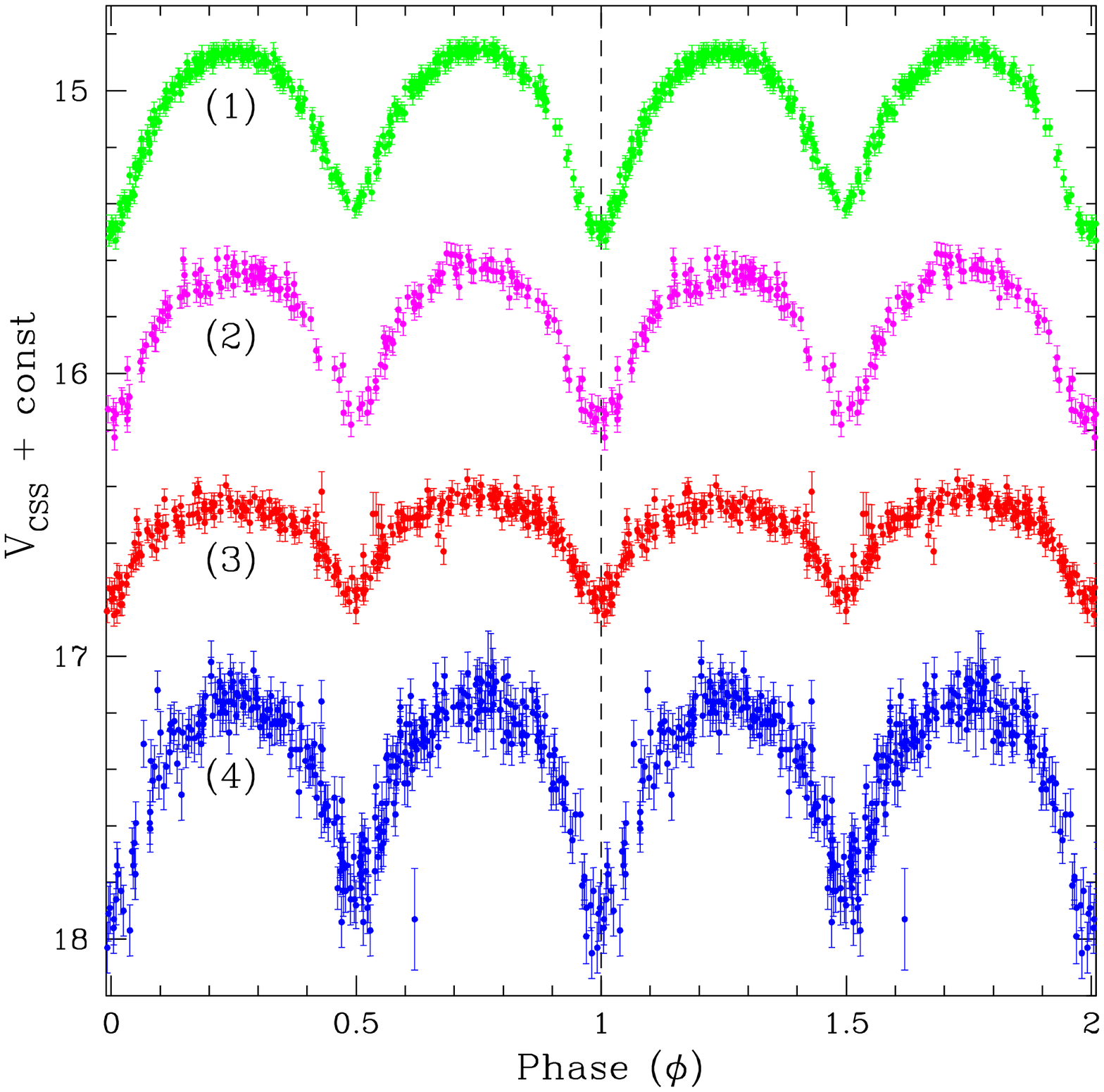}{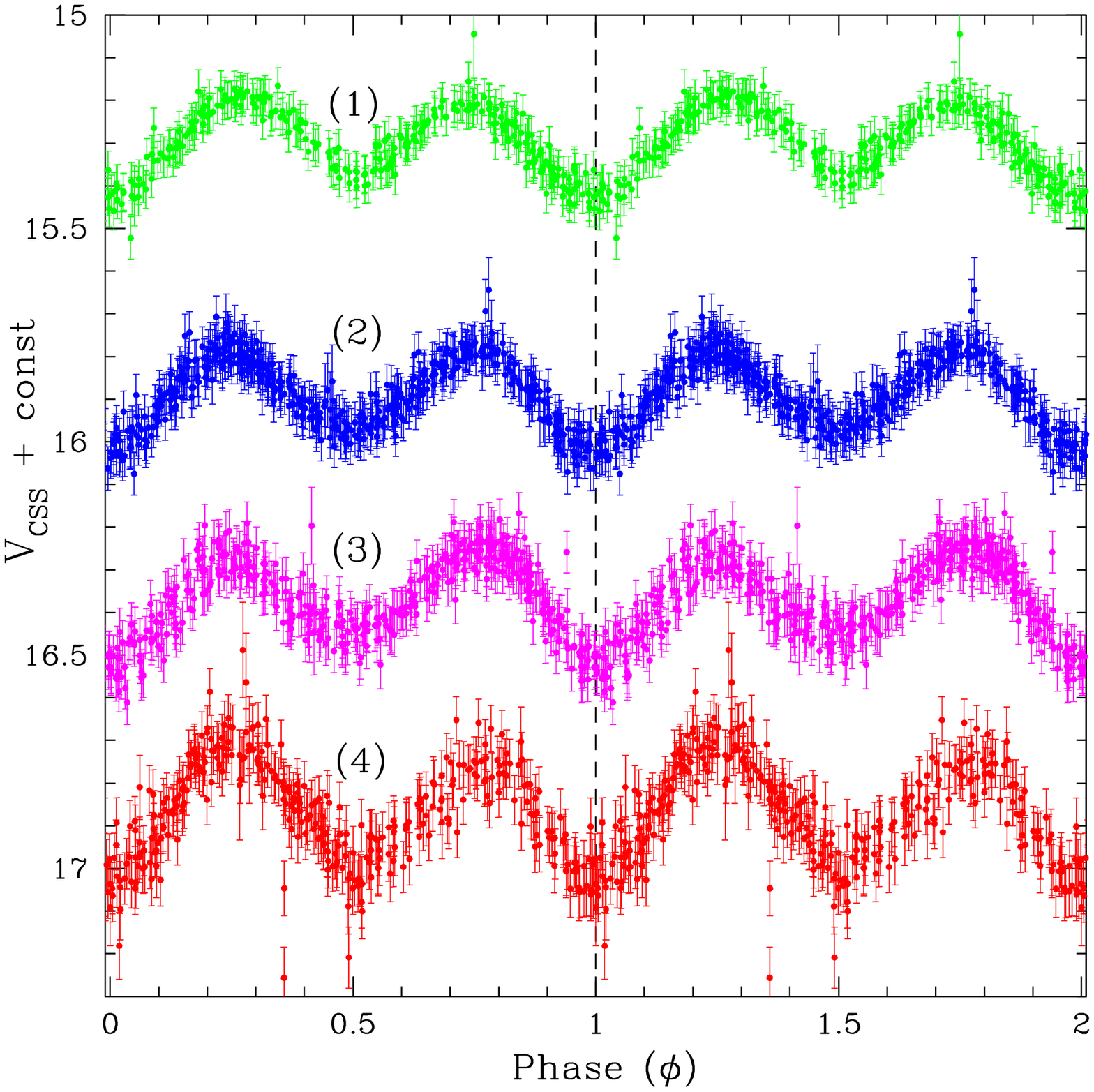}
\caption{\label{STypes}
Examples of ultra-short period binary light curves.
In the left panel, the plots labeled 1, 2, 3 and 4 are contact binaries 
CSS\_J041950.2-012612 (green, $P=0.196$ days), 
CSS\_J112237.1+395219 (magenta, $P=0.197$ days),
CSS\_J112243.5+372130 (red, $P=0.169$ days),
and CSS\_J140027.3+060748 (blue, $P=0.192$ days), 
respectively. 
In the right panel,  the plots labeled 1, 2, 3 and 4 are ellipsoidal binary candidates 
CSS\_J001242.4+130809 (green, $P=0.164$ days), CSS\_J081158.6+311959 (blue, $P= 0.156$ days),  
CSS\_J090119.2+114254 (magenta, $P=0.187$ days), and CSS\_J111647.8+294603 (red, $P=0.146$ days), 
respectively.
}
}
\end{figure*}

Although over-contact, contact, semi-detached, and detached binaries can exhibit quite distinct light curve shapes, they
appear similar at the limit between these types of binary systems. Or alternatively, when they are observed with low
signal-to-noise.  However, in this analysis we found that the light curves for short period binaries were generally
quite well separated into just two groups.  In Figure \ref{STypes}, we contrast the two main types of short period light
curves.

This same shape division can also be seen in the light curves of the short period binary candidates presented
by Nefs et al.~(2012). However, the light curves of ellipsoidal variables closely resemble those of over-contact
binaries. For ellipsoidal variables, the variation is due to the distorted shapes of a star or stars.  This can occur
even when the objects are far from contact.  Thus the sinusoidal light curves of some of the Nefs et al.~(2012)
short period contact binary candidates are consistent with both over-contact binaries and the ellipsoidal variations
due to compact companions such white dwarfs (WDs) or subdwarfs.  Such short-period sinusoidal light curves have been found
in many white dwarf main sequence (WDMS) systems (e.g.~Rodriguez-Gil et al.~2009, Pyrzas et al.~2012, Parsons et
al.~2013). In fact, systems with such light curves were specifically rejected from the short period main sequence contact
binary sample of Norton et al.~(2011), because of possible contamination of WDMS systems.

For ellipsoidal variables, the amplitude is generally limited to $\sim 0.3$ mag, since the variations are due to the
distorted star filling its Roche lobe (Parsons et al.~2013). At very short periods, a distorted star will overflow its
Roche lobe becoming a cataclysmic variable. The accretion induced variability in CV systems generally distinguishes them
from stable non-accreting WDMS systems.

\begin{figure}[ht]{
\epsscale{1.0}
\plotone{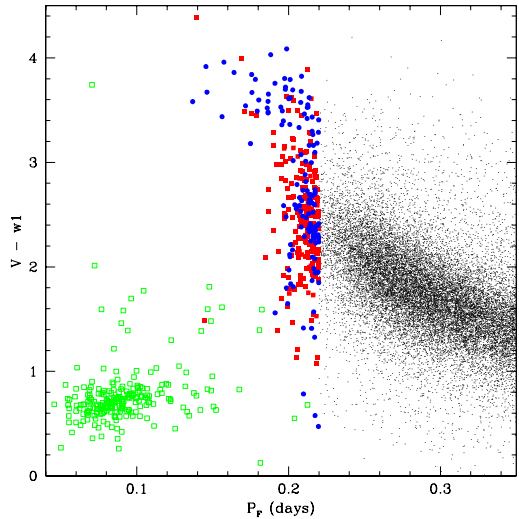}
\caption{\label{WiseS}
Period-colour distribution for short period variables.
$\delta$ Scuti variables (green open boxes), ellipsoidal candidates
(blue dots), short period eclipsing binaries (red boxes), 
general ellipsoidal and eclipsing binaries (black points).
}
}
\end{figure}

To investigate the nature of the objects with short period objects, we looked at the colours of sources in the two
groups using WISE, SDSS and GALEX data.  In Figure \ref{WiseS}, we show the WISE data for the eclipsing binaries as well
as $\delta$ Scuti stars (that have similar periods). The separation between the pulsating $\delta$ Scuti stars and the
eclipsing and ellipsoidal-like ones is quite clear. However, some of the ellipsoidal candidates do have colours similar to
$\delta$ Scutis.  These objects may either be systems where the companion is a hot WD, or incorrectly classified
$\delta$ Scutis. There is clearly significant overlap between ellipsoidal candidates and eclipsing systems. There are
also ellipsoidal candidates that have $V - w1 > 3.5$. These objects are redder than most of the eclipsing objects and
thus suggest late stellar types.

\begin{figure*}[ht]{
\epsscale{1.0}
\plottwo{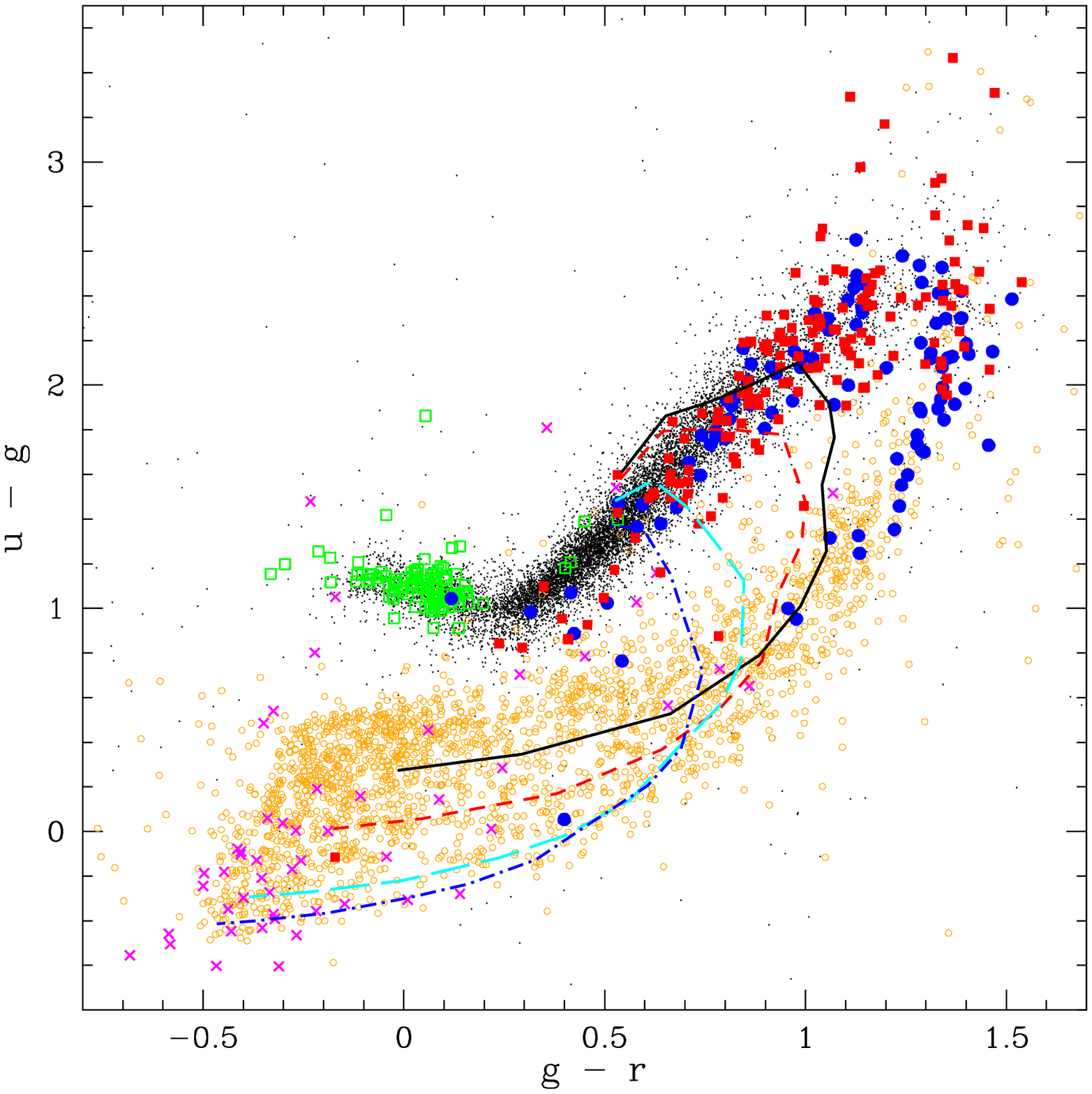}{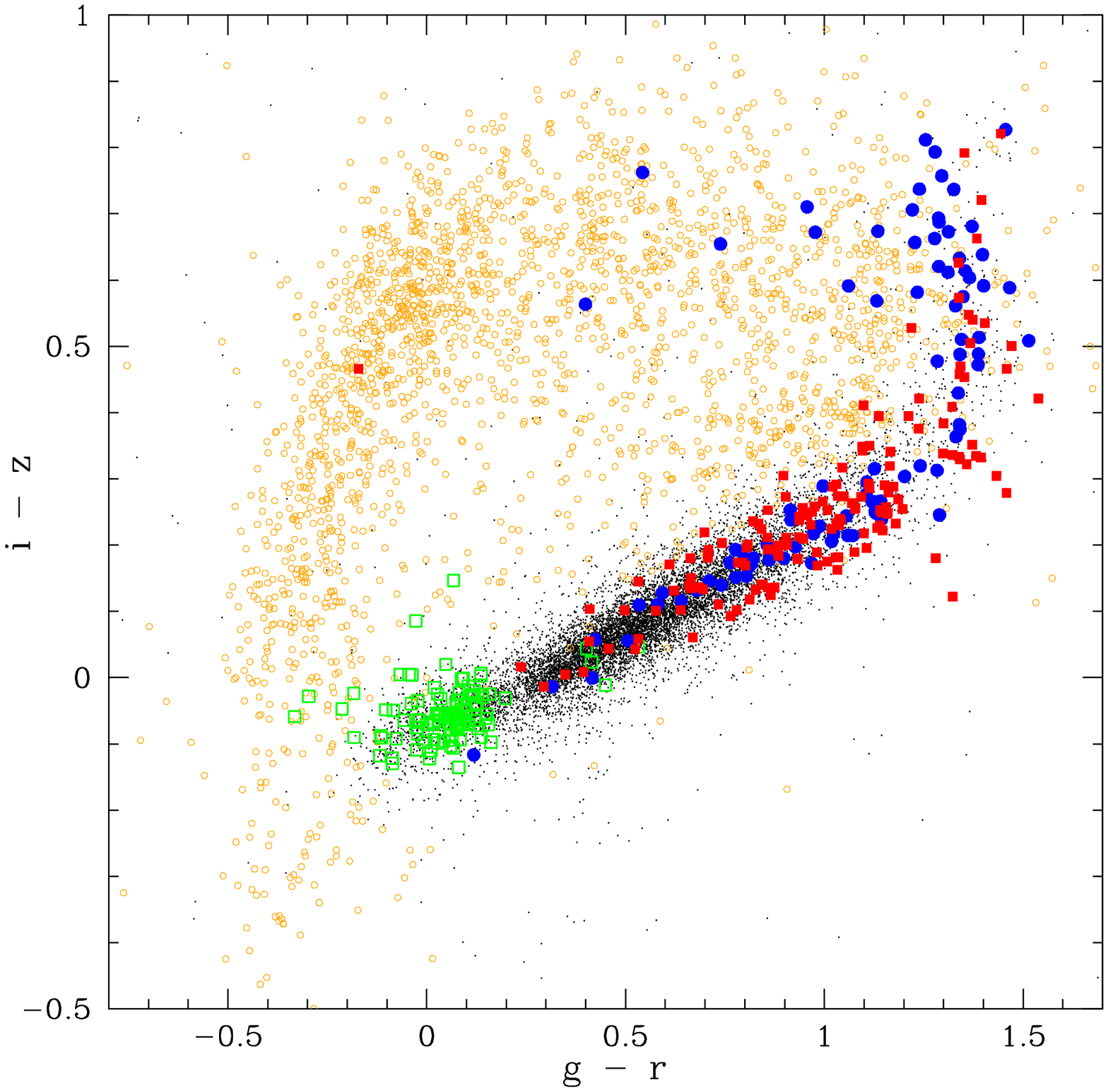}
\caption{\label{SDSSS}
Optical colour distribution for short period variables. Colours based on extinction 
correct SDSS photometry. The symbols are the same as Figure \ref{WiseS} with the 
addition of known CVs and white dwarfs shown as magenta crosses and spectroscopically 
selected WD+dM binaries from Rebassa-Mansergas et al.~(2012) as small orange circles. 
The solid black line, short-dashed red line, long-dashed cyan line and blue dash-dotted
line give model colours of WDMS binaries where the WDs have effective temperatures of 
10,000K, 20,000K, 30,000K and 40,000K, respectively from Rebassa-Mansergas et al.~(2012).
}
}
\end{figure*}

In Figure \ref{SDSSS}, we show the $u-g$ and $i-z$ vs $g-r$ plots for the contact eclipsing binaries from Drake et
al.~(2014) with SDSS DR9 photometry along with $\delta$ Scuti stars and spectroscopically identified WD+dM binaries from
the Rebassa-Mangergas et al.~(2012) catalog. We also plot the WD+dM model isochrones given by Rebassa-Mangergas et
al.~(2012) and colours of other known WDMS systems detected in CSDR1 by Drake et al.~(2014).

Figure \ref{SDSSS} also shows that many of the short period ellipsoidal candidates, exhibit excess colour in $g-r$
compared to main sequence eclipsing binaries. This excess is completely consistent with the colours of spectroscopically
identified WD+dM binaries from Rebassa-Mansergas et al.~(2012). However, the colours of most objects lie at the red end of
the known distribution, where WDs cooler than 10,000K are expected. Such systems are very difficult to identify
spectroscopically since cool WDs can be much fainter than M-dwarfs. Thus they do not give rise to significant $u$-band flux.
From $i-z$ we see that the ellipsoidal candidates are generally redder that the regular eclipsing sources, suggesting
that the M-dwarfs in WDMS pairs are usually later types than those in MS binary pairs.  The excess $g-r$ colour is
indicative of a warm secondary, rather than a cool M-dwarf.  In contrast, we see that most of the objects with regular eclipsing
lightcurves do not exhibit any $g-r$ colour excess, suggesting that most are truly main sequence pairs. Although there
may still be some confusion (as outlined below). 

Aside from the true colour features, we note that the u-band magnitudes of the redder candidates are likely to be
affected to some extent by the SDSS red leak\footnote{http://www.sdss3.org/dr9/imaging/caveats.php}.  There is 
also likely to be a measurable amount of scatter in colour due to the the SDSS data being taken at a single epoch.

In cases where one of the components is a hot WD, one expects the presence of significant UV flux. Therefore, we matched
the periodic variables with GALEX sources (within $5\arcsec$). We found matches for 20,000 of the 61,000 sources in Drake
et al.~(2014) periodic variable catalog.

\begin{figure*}[ht]{
\epsscale{1.0}
\plottwo{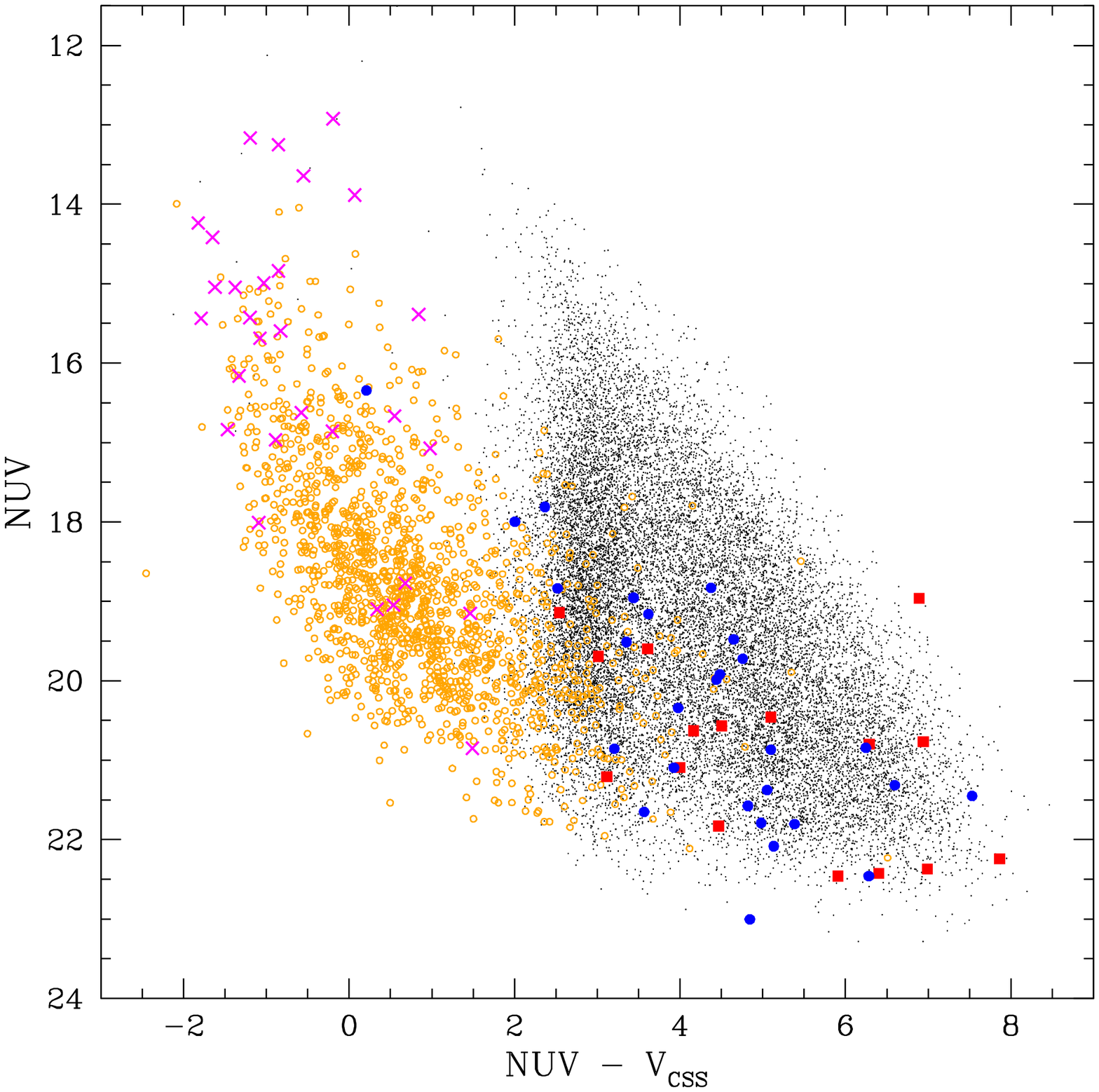}{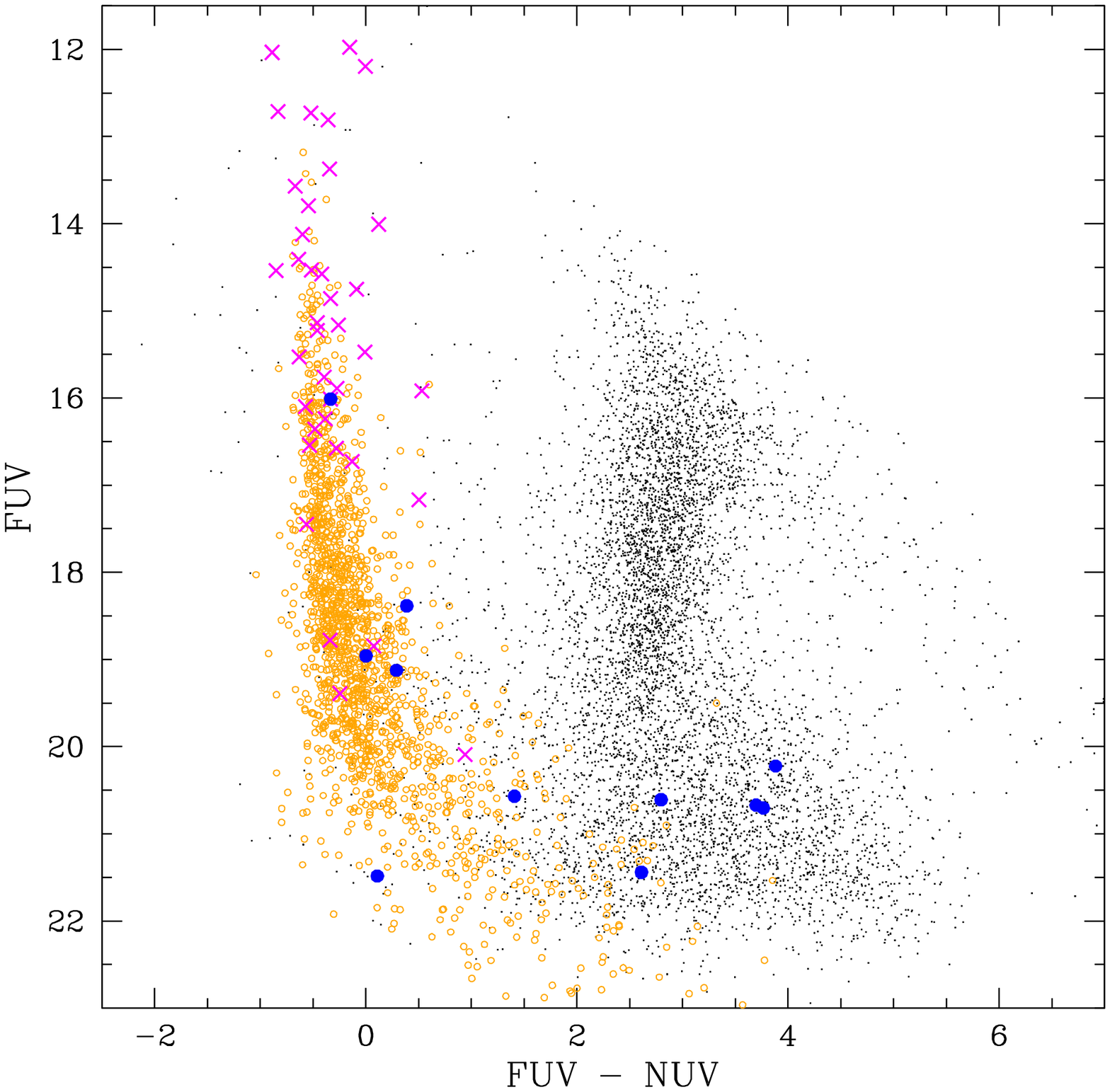}
\caption{\label{GalS}
UV and optical colours of the periodic variable stars.  In the left panel we plot the GALEX GR5 near-UV colours of the
periodic variables. The symbols match those given in Figure \ref{SDSSS}.
}
}
\end{figure*}

In Figure \ref{GalS}, we compare the ellipsoidal candidates and eclipsing sources, with the known WD binaries as noted
earlier.  The eclipsing and ellipsoidal systems are not well separated in $NUV-V_{CSS}$ colour and overlap with other
periodic variables (as do many of the spectroscopically WDMS systems). Some of the ellipsoidal candidates do stand out
in $FUV-NUV$ colours. However, very few are detected at the depth of GALEX $FUV$ measurements. As expected, the 
short-period eclipsing systems are not detected in GALEX FUV data.

The GALEX results strongly suggest that the companions must be much cooler than most of the known WDMS systems.  Since
WDs are among the most common stars, it is expected that numbers of cool WD+dM binaries should exceed the number of
hot systems (Rebassa-Mansergas et al.~2013).  Low accretion-rate CVs and pre-CV systems, such as SDSS 121010.1+334722.9
(Pyrzas et al.~2012), may also produce very little emission with cool WDs.  The detection of FUV emission from the ellipsoidal
candidates strongly suggests that some fraction of these systems contain hot WDs.

\begin{figure}[ht]{
\epsscale{1.0}
\plotone{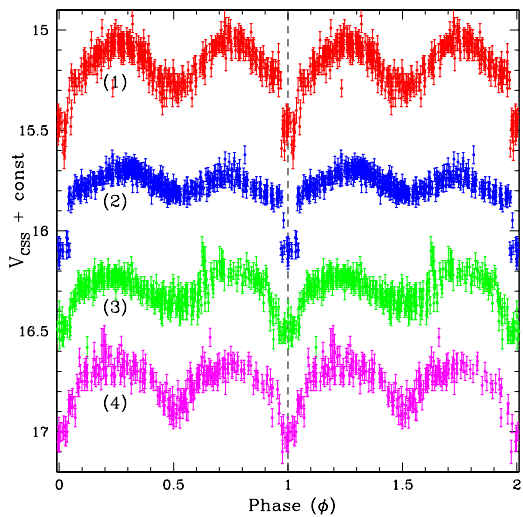}
\caption{\label{Ell_EclS}
Candidate ellipsoidal variables with discrete eclipsing features (top three curves) compared to
a morphologically similar, likely contact binary. The variables labeled 1 through 4 
are CSS\_J090826.3+123648 (red, $P=0.139$ days), 
CSS\_J093508.0+270049 (blue, $P=0.201$ days), 
CSS\_J165352.4+391410 (green, $P=0.220$ days), 
and CSS\_J234131.5+375439 (magenta, $P=0.171$ days), 
respectively.
}
}
\end{figure}

Another piece of evidence suggesting that ellipsoidal candidates are not contact binaries comes from the fact that some
of the light curves exhibit clear eclipses. In Figure \ref{Ell_EclS}, we plot examples of short-period systems where a
compact source eclipses a distorted companion. The shallow depth of the eclipses, along with the short ingress and
egress times, suggests that the eclipsing source is much more compact than the distorted companion.  The light curves
also lack any obvious secondary eclipse. This further suggests that compact eclipsing source is much fainter than the star
undergoing the eclipse. Thus in these cases at least, one of the components must be fainter and smaller than their
companions.

The lower two light curves of Figure \ref{Ell_EclS} demonstrate the difficulty there can be when discerning whether a system 
has the W-shaped eclipses of a contact binary, or the sinusoidal light curve of an ellipsoidally distorted M-dwarf with a
faint, compact companion.

For most ellipsoidal candidates there was no sign of distinct eclipses. It is possible that some of the companions could
be neutron stars.  Indeed, $\gamma$-ray pulsar PSR J2339-0533 (Romani \& Shaw 2011) was among the periodic sources
detected in our analysis of periodic variables in CSDR1. This pulsar exhibits a high level of variability due to a
highly distorted companion. It also has a period in the range of these sources (0.193 days). We matched the short-period
sources with unidentified sources in the Fermi-LAT catalog (Nolan et al.~2012) but found no convincing matches. In this
short-period range we expect that low mass companions to neutron stars should give rise to a higher level of variability
than observed, suggesting that most of systems are likely to be due to WDs rather than other compact sources.

\subsection{Spectroscopic Follow-up}

In order to confirm the nature of the ultra-short period binary candidates we carried out low-resolution spectroscopy
with the Optical System for Imaging and Low Resolution Integrated Spectroscopy (OSIRIS) tunable imager and spectrograph
(Cepa et~al. 2003; Cepa 2010) at the 10.4\,m Gran Telescopio CANARIAS (GTC), located at the Observatorio Roque de los
Muchachos in La Palma, Canary Islands, Spain.  The heart of OSIRIS is a mosaic of two 4k\,$\times$\,2k e2v CCD44--82
detectors that gives an unvignetted field of view of 7.8\,$\times$\,7.8\,arcmin$^{2}$ with a plate scale of
0.127\,arcsec\,pix$^{-1}$.  However, to increase the signal-to-noise ratio of our observations, we chose the standard
operation mode of the instrument, which is a 2\,$\times$\,2-binning mode with a readout speed of 100\,kHz. All spectra
were obtained with the OSIRIS R1000B grism. We used the 1.23\,arcsec-width slit, oriented at the parallactic angle to
minimize losses due to atmospheric dispersion.  The resulting resolution, measured on arc lines, was R $\sim$ 700 in the
approximate 3500--8000\,{\AA} spectral range.  An additional seven spectra were obtained with the Palomar 5m telescope
(P200) using the Double Beam Spectrograph (DBSP) using a 1\,arcsec-width slit and spectral range 3700--10000\,{\AA}. 

All the spectra were reduced using standard reduction procedures within IRAF. The GTC observing program concentrated on
taking spectra of the short-period objects with contact binary light curves, while the Palomar program concentrated on
sources with the shortest periods (which are mostly ellipsoidal candidates).  In addition to these programs, we matched
all the ultra-short period binaries with objects having spectra within the Sloan Digital Sky Survey Data Release 10
(SDSS DR10; Ahn et al.~2014).  We found three SDSS matches to our eclipsing candidates, and one to an ellipsoidal
candidate.

\begin{table*}
\scriptsize
\caption{Spectra of Ultra-short Period Binaries}
\label{TabSpec}
\begin{center}
\begin{tabular}{@{}lllllcccccc}
\hline
CRTS ID & RA & Dec (J2000) & $V_{CSS}$ & Period & Telescope & Category & $u-g$ & $g-r$ & Spectral Type\\
\hline
CSS\_J090826.3+123648 & 09:08:26.26 & +12:36:49.0 & 15.27 & 0.1391985 & P200 & EA/ELL & 0.87 & 0.78 &M7V\tablenotemark{a,b}\\
CSS\_J111647.8+294602 & 11:16:47.83 & +29:46:02.8  & 17.06 & 0.146249 & P200/SDSS & ELL & 1.73 & 1.27 & M3.5V\tablenotemark{b,c}\\
CSS\_J081158.6+311959 & 08:11:58.58 & +31:19:59.5 & 16.13 & 0.156187 & P200 & ELL & 2.39 & 1.51 & M2V\tablenotemark{b}\\
CSS\_J001242.4+130809 & 00:12:42.41 & +13:08:09.6 & 15.48 & 0.164086 & P200 & ELL & 1.78 & 1.29 & M5V\tablenotemark{b}\\
CSS\_J112243.5+372130 & 11:22:43.44 & +37:21:30.2 & 16.56 & 0.168744 & P200 & EW  & 2.70 & 1.44 & M4.5V\\
CSS\_J041950.2-012612 & 04:19:50.16 & $-$01:26:12.1 & 17.87 & 0.175752 & P200 & EW & \nodata & \nodata & M3V\\
CSS\_J171508.5+350658 & 17:15:08.50 & +35:06:58.7 & 16.56 & 0.178549 & P200/GTC & EW & 2.75 & 1.40 & M2V\\
CSS\_J112237.0+395219 & 11:22:37.06 & +39:52:19.9  & 17.07 & 0.184749  & SDSS & EW & 1.51 & 0.62 & F9V\\
CSS\_J090119.2+114254 & 09:01:19.22 & +11:42:54.7 & 16.34 & 0.1866878 & P200 & ELL & 0.95 & 0.98 & M4V\tablenotemark{b}\\
CSS\_J151531.4-153418 & 15:15:31.43 & $-$15:34:18.6  & 17.61 & 0.190148 & GTC  & EW  & \nodata & \nodata & K5V\\ 
CSS\_J101022.0+012438 & 10:10:22.01 & +01:24:38.5 &  18.8 & 0.190321 & P200 & EW  & 2.34  & 1.38 & M0V\\
CSS\_J013224.5+011522 & 01:32:24.53 & +01:15:22.0  & 15.04 & 0.196014 & GTC  & EW  & \nodata & \nodata & K9V\\ 
CSS\_J161945.9+241312 & 16:19:45.93 & +24:13:12.4  & 17.86 & 0.19727  & GTC  & EW  & 1.91 & 1.04 & K6V\\ 
CSS\_J044647.5+021635 & 04:46:47.55 & +02:16:35.0  & 18.03 & 0.19777  & GTC  & EW  & \nodata & \nodata &  K5V\\ 
CSS\_J234002.2+204122 & 23:40:02.25 & +20:41:22.2  & 16.29 & 0.198047 & GTC  & EW  & 2.76 & 1.33 & K7V\\ 
CSS\_J213019.2-065136 & 21:30:19.20 & $-$06:51:36.2  & 16.42 & 0.198785 & GTC  & EW & 2.24 & 1.38 & M2V\\ 
CSS\_J115533.4+354439 & 11:55:33.44 & +35:44:39.3  & 15.77 & 0.199725 & GTC  & EW  & 2.11 & 1.34 & M2V\\ 
CSS\_J012119.1-001950 & 01:21:19.12 & $-$00:19:50.7  & 15.70 & 0.207282 & SDSS & EW & 2.52 & 1.21 & K7V\\
CSS\_J122814.6+534746 & 12:28:14.61 & +53:47:46.7  & 17.23 & 0.212048 & SDSS & EW  & 2.25 & 1.08 & K5V\\
\hline
\tablenotetext{a}{Matches high proper motions system LP 486-53, with $\Delta \alpha = -122$, $\Delta\delta= -172$ mas/yr (Lepine \& Shara 2005).}
\tablenotetext{b}{System where radial velocity variations were measured.}
\tablenotetext{c}{Matches SDSS J111647.81+294602.7. Previously identified as candidate binary based 
on SDSS spectra by Clark, Blake \& Knapp (2012).}
\end{tabular}
\end{center}
Col. (1) Catalina ID.
Cols. (2) \& (3) Right Ascension and Declination
Col. (4) Period in days.
Col. (5) Source of the spectrum where P200 is the Palomar 5m, and GTC is Gran Telescopio Canarias 10.4m.
Col. (6) Type of periodic variable (ELL = ellipsoidal candidate, EW = contact or W UMa type, EA = detached or Algol type).
Cols. (7) \& (8) Extinction corrected colours from SDSS DR10 photometry.
Col. (9) Spectral type.
\end{table*}

\begin{figure*}[ht]{
\epsscale{1.0}
\plottwo{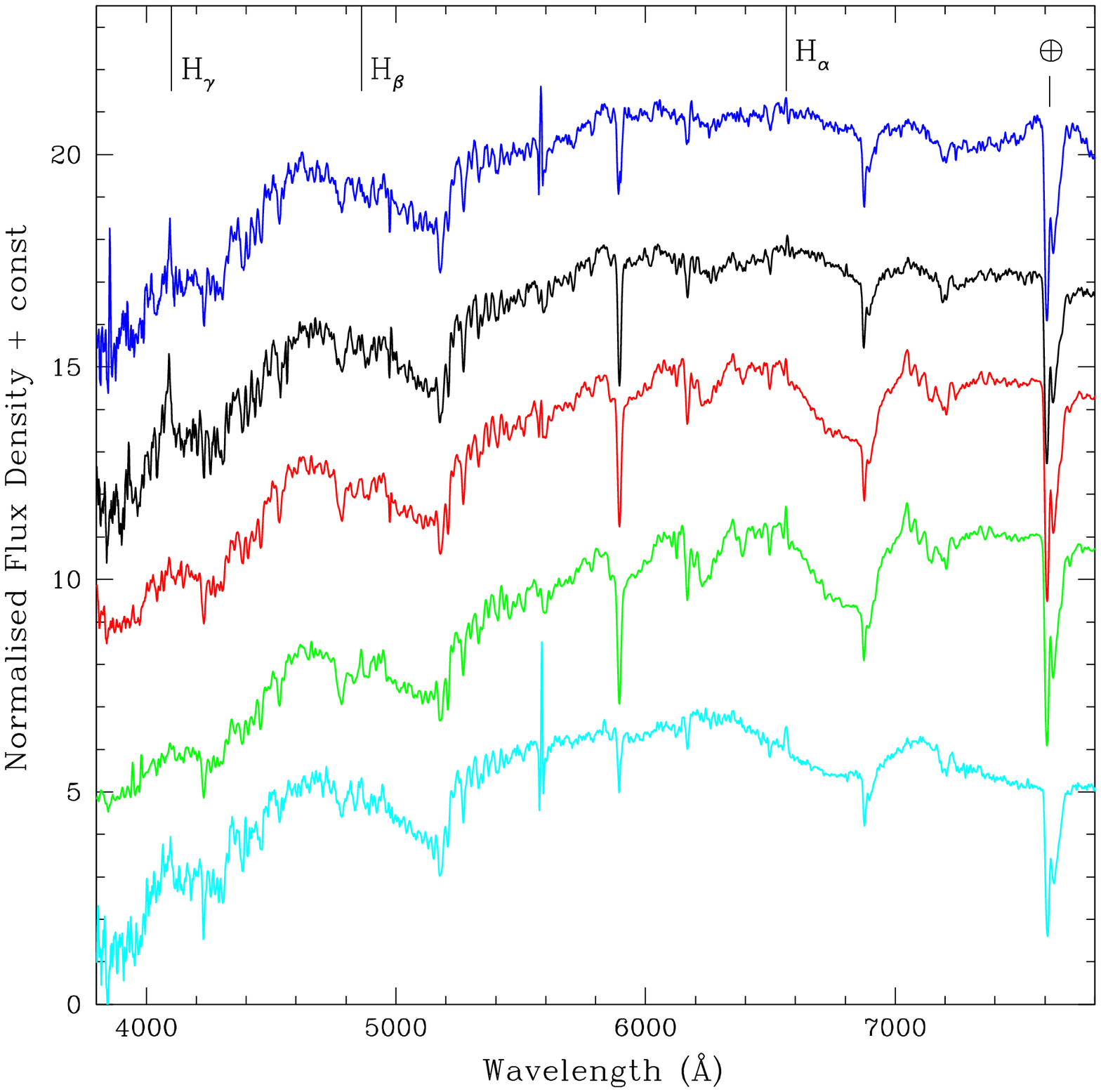}{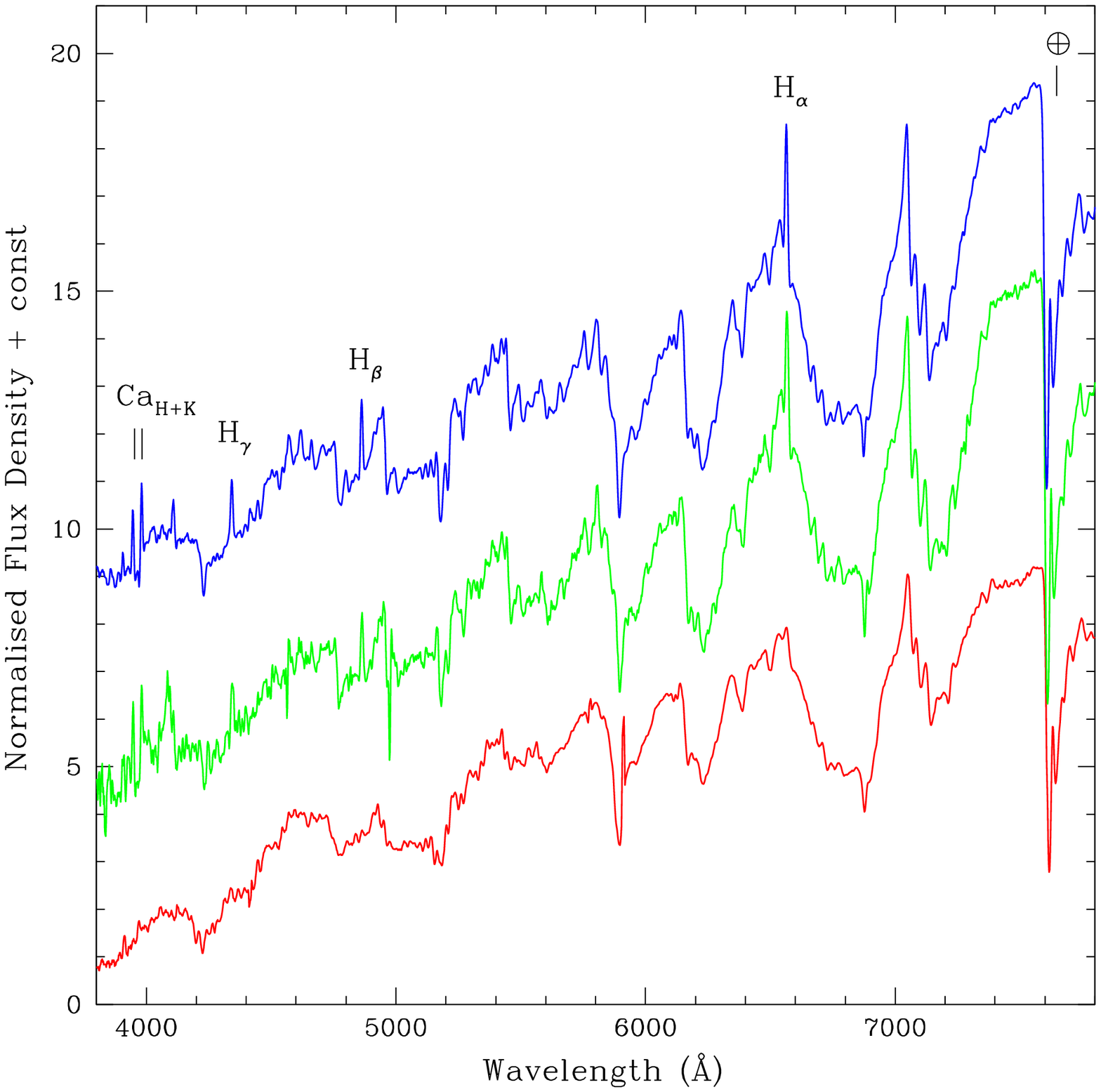}
\caption{\label{GTCspec}
GTC optical spectra of ultra-short period eclipsing
binary candidates. In the left panel (from top to bottom), 
we plot the spectra for contact binaries with K-types:
CSS\_J151531.4-153418 (blue); CSS\_J044647.5+021635 (black);
CSS\_J234002.2+204122 (red); CSS\_J013224.5+011522 (green)
and CSS\_J161945.9+241312 (cyan).
In the right panel (from top to bottom), we plot the spectra for 
sources with M spectral types: CSS\_J115533.4+354439 (blue); CSS\_J213019.2-065136 (green) 
and and CSS\_J171508.5+350658 (red).
}
}
\end{figure*}

\begin{figure*}[ht]{
\epsscale{1.0}
\plottwo{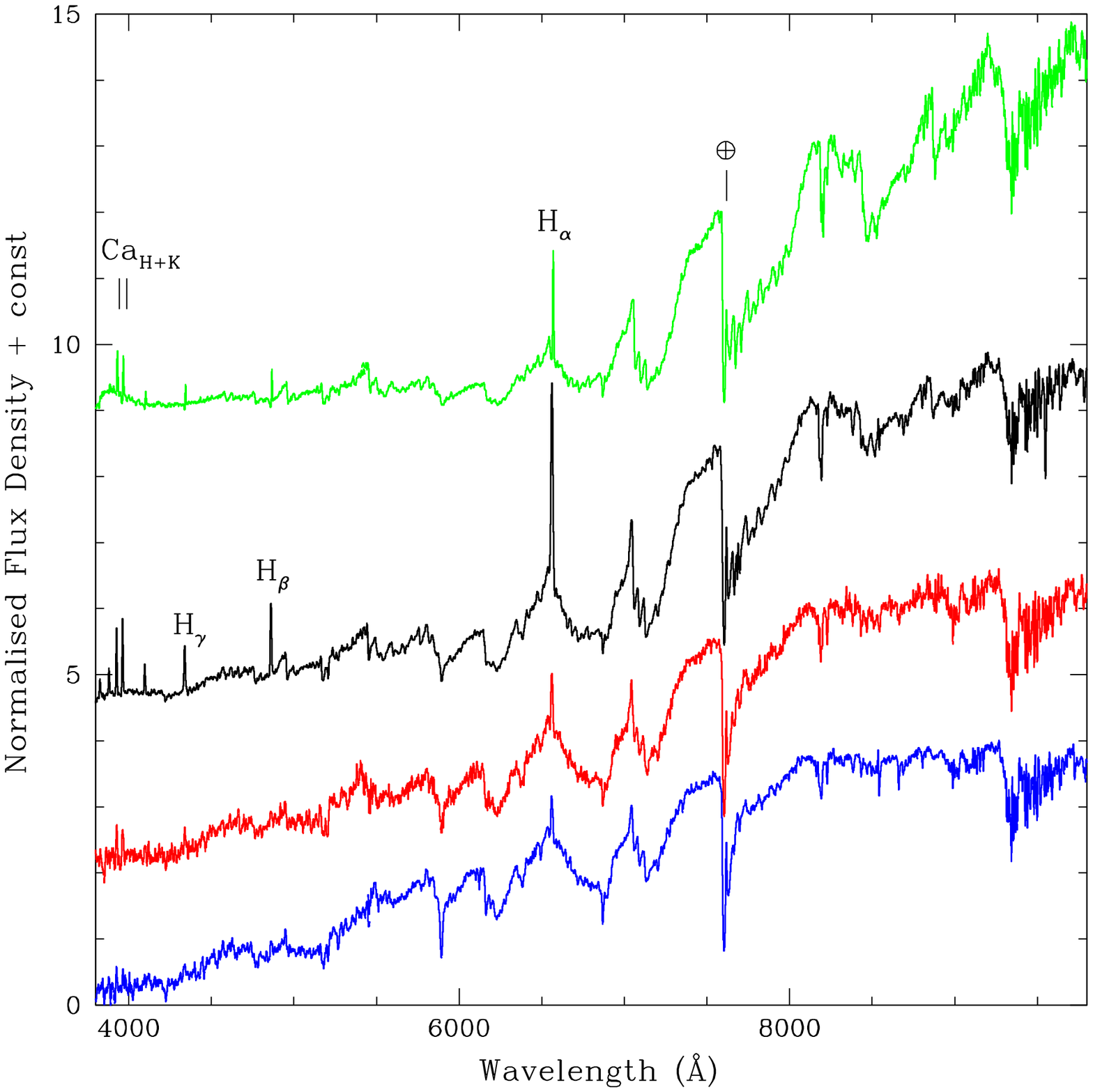}{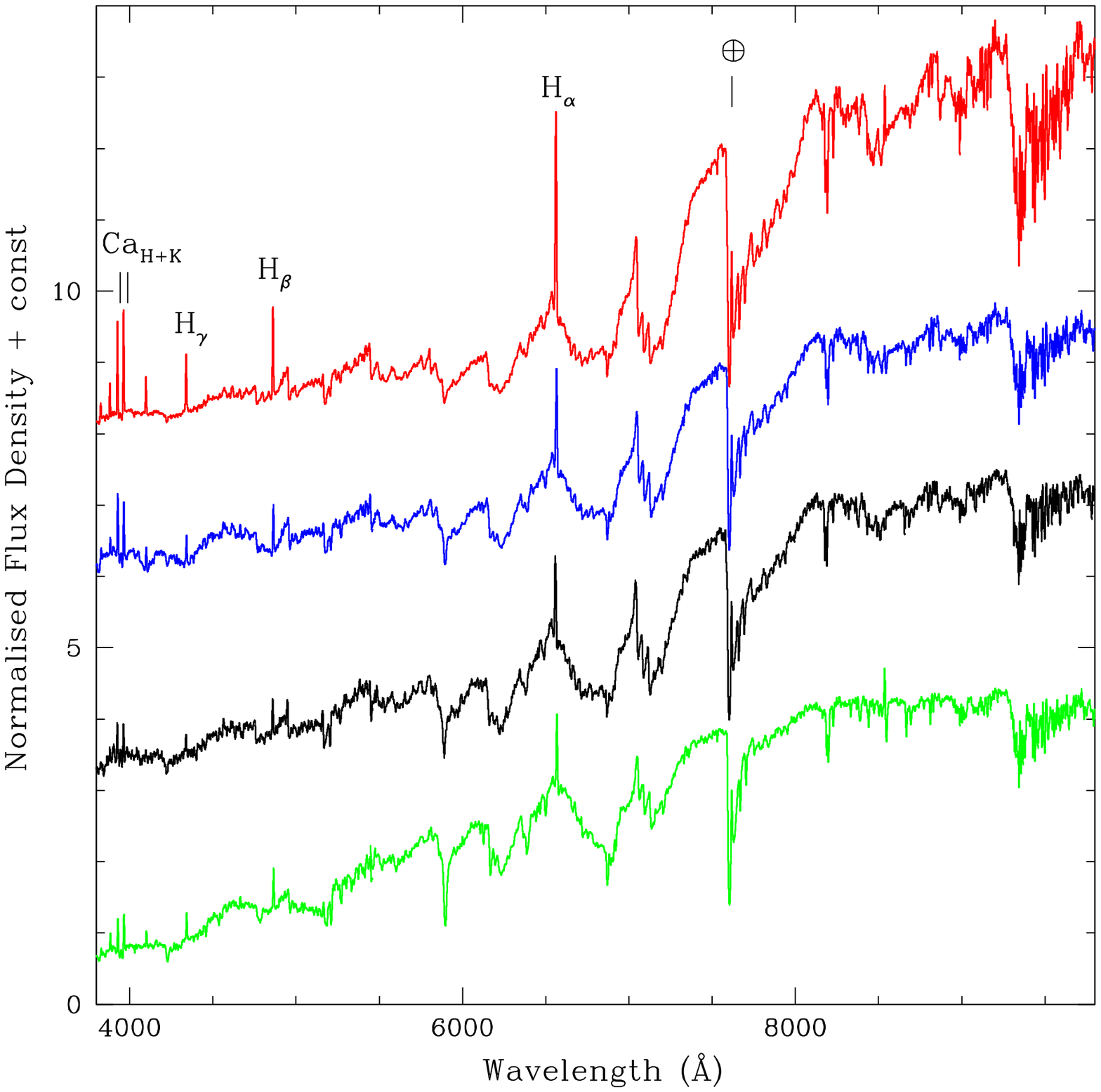}
\caption{\label{P200spec}
Palomar optical spectra of ultra-short period eclipsing
binary candidates. In the left panel (from top to bottom), 
we plot the contact binaries sources:
CSS\_J090826.3+123648 (LP 486-53, green);
CSS\_J112243.5+372130 (black);
CSS\_J041950.2-012612 (red),
and CSS\_J171508.5+350658 (blue).
In the right panel (from top to bottom), we plot the ellipsoidal candidates:
CSS\_J001242.4+130809 (red);
CSS\_J090119.2+114254 (blue);
CSS\_J111647.8+294603 (SDSS J111647.81+294602.7, black)
and CSS\_J081158.6+311959 (green).
}
}
\end{figure*}

\begin{figure}[ht]{
\epsscale{1.0}
\plotone{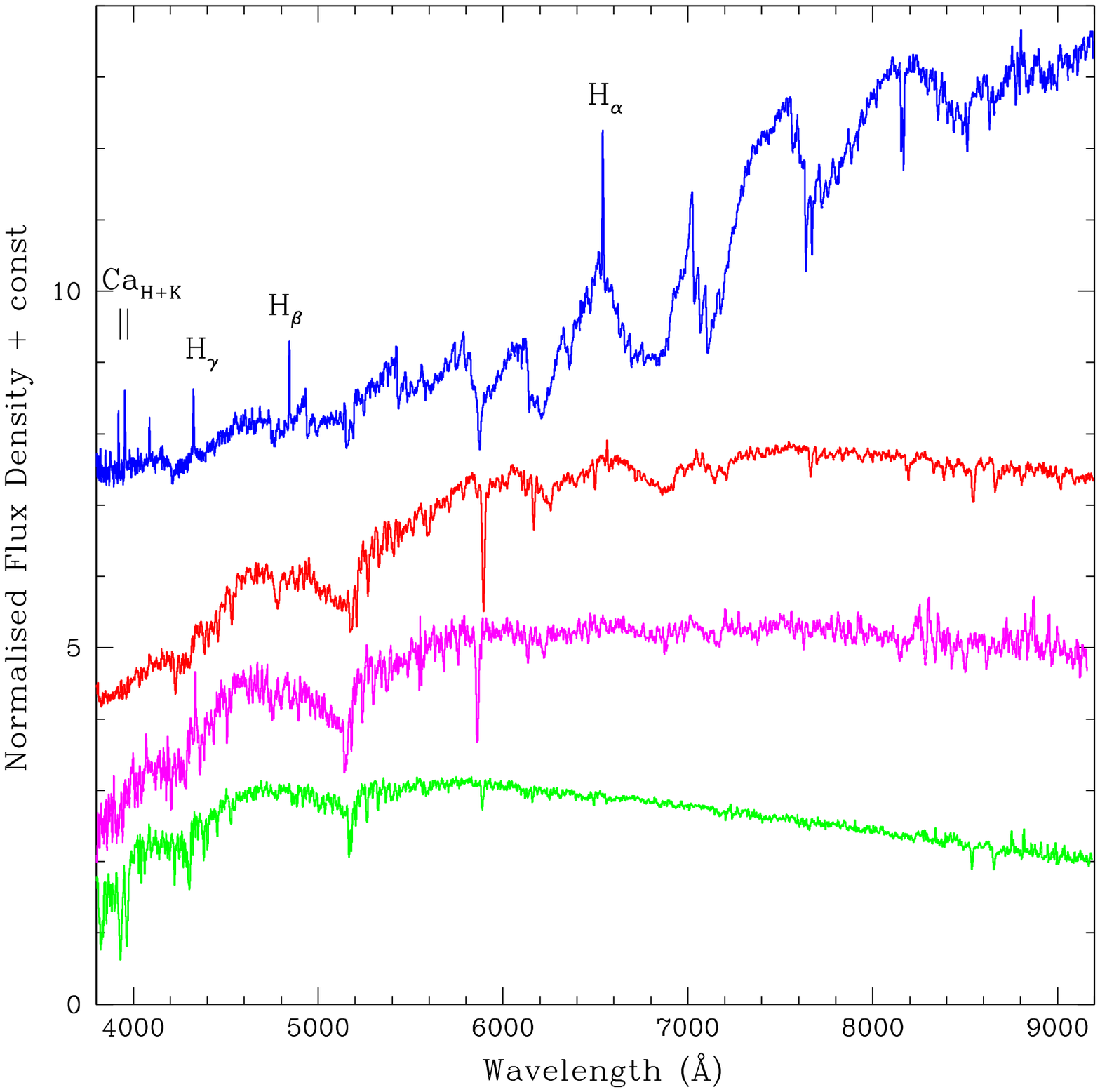}
\caption{\label{SDSSspec}
SDSS optical spectra of ultra-short period eclipsing binary candidates. 
These spectra include (from top to bottom):
CSS\_J111647.8+294603 (blue); 
CSS\_J112237.0+395219 (red); 
CSS\_J012119.1-001950 (magenta) 
and CSS\_J122814.6+534746 (green). 
}
}
\end{figure}

In Table \ref{TabSpec}, we present the details of the SDSS, Palomar and GTC spectra and in Figures \ref{GTCspec},
\ref{P200spec} and \ref{SDSSspec}, we present the GTC, Palomar and SDSS spectra, respectively.  Spectral types were
derived by comparison to SDSS classifications. In particular, the M-dwarfs were classified using the high
signal-to-noise combined templates of Bochnaski et al.~(2007). Overall, the GTC spectra consist of five K-dwarfs and
three early M-dwarfs, while the shorter period systems observed with Palomar all contain M-dwarfs. One system with an
SDSS spectrum, and one object observed by GTC, were re-observed with Palomar as a consistency check. The
spectral types were in excellent agreement.

Inspection of the spectra revealed the presence of clear Balmer emission in almost all of the spectra. Such emission is
seen in a large fraction of M-dwarfs and increases significantly for late types (Bochnaski et al.~2005).  However, only
$\sim 10\%$ of M1 and M2 stars are known to exhibit such activity.  Furthermore, balmer emission is also seen in the 
K-dwarfs observed by GTC. Only a couple of percent of the stars are expected to exhibit emission (Zhao et al. 2013).
The enhanced number here is a sign of magnetic activity induced in these compact binaries. Similar highly 
enhanced active fractions were found WD+dM systems by Morgan et al.~(2012). Additionally, Ca H+K lines are 
present in emission in most of the spectra. This is also a sign of strong magnetic interaction between the components. 

Comparison of the spectra for the ellipsoidal candidates with the regular eclipsing binaries, does not show any obvious
difference. Among the sources from Table \ref{TabSpec}, only CSS\_J090119.2+114254, and CSS\_J090826.3+123648, show some
evidence for an additional component at the bluest wavelengths.  As noted in Table \ref{TabSpec}, these two stars also
have the bluest colours ($u-g < 1$ and $g-r < 1$) of the candidates with SDSS photometry.  The lack of clear evidence
for a white dwarf companion in the ellipsoidal systems is not unexpected since van den Besselaar et al.~(2007),
Rodriguez-Gil et al.~(2009), Rebassa-Mansergas et al.~(2010), Pyrzas et al.~(2012), Parsons et al.~(2013) and
Rebassa-Mansergas et al.~(2013) all present examples of WDMS spectra where evidence for the WD is not clearly
discerned.  The lack of a blue component in these spectra suggests that the sources causing the distortion are generally
cooler than the hot WDs that have previously been found using SDSS spectra (Rebassa-Mansergas et al.~2012). 
The absence of a blue component is also expected based on the SDSS and GALEX colours presented earlier.

\subsubsection{Radial Velocities}

Since the spectra of the short-period systems poorly constrain cases where the primary is too faint to be seen, we
undertook a program to better classify the binary systems using radial velocity variations.  For short-period systems,
an unseen yet relatively massive companion (such as a WD), is expected to produce large radial velocity variations
in low mass companion stars (such as M-dwarfs). 
The presence of small velocity variations can also be used to exclude possible sources of contamination, such as
pulsating stars. For example, $\delta$ Scuti variables have velocity variations of only a few km/s (Penfold 1971, Zima
et al.~2007, Antoci et al.~2013), while WD+dM binaries have typical radial velocity amplitudes of around 150km/s (Nebot
G\'omez-Mor\'an et al. 2011).

Multiple epochs of optical spectra were obtained for five of the systems noted in Table \ref{TabSpec}. As the spectra
only exhibit clear emission from one component, it is only possible to measure the velocity variations of a single
source in each system. Future high resolution observations of these objects may enable the detection features from both
sources.

For each source, we measured velocity variations by the average Doppler velocities of the Balmer and Ca II H+K lines.  In
each case the average was weighted by the relative strength of the lines.  The velocity errors are for these low
resolution spectra are of order $\sim 15km/s$.  As the light curves themselves are measured over a period of years
the phases of our observations are expected to accurate to $\sim 0.01$ cycles.

\begin{table*}
\caption{Ultra-short Period Binary fits}
\label{TabFit}
\begin{center}
\begin{tabular}{@{}crclllllll}
\hline
CRTS ID & $\rm V1_{max}$  & $\rm V2_{max}$  & M1  & M2   & Teff1 & Teff2  & I  &  R1  & R2\\
& (km/s) & (km/s) & ($\rm M_{\sun}$) & ($\rm M_{\sun}$) & (K) & (K) & (deg) & ($\rm R_{\sun}$) & ($\rm R_{\sun}$)\\
\hline
CSS\_J001242.4+130809 &  77   & 110  & 0.24 & 0.17 & 3200 & 2900 & 40.5  & 0.410 & 0.36\tablenotemark{a}\\ 
  '   '   '   & 101   & 225  & 0.58 & 0.22 & 6100 & 2900 & 70.9  & 0.034 & 0.34\tablenotemark{b}\\
CSS\_J090119.2+114254 &  54   & 256  & 0.58 & 0.12 & 7800 & 3100 & 69.9  & 0.018 & 0.58\\
CSS\_J090826.3+123648 &  86   & 294  & 0.62 & 0.18 & 6500 & 2500 & 85.9  & 0.008 & 0.26\\ 
CSS\_J111647.8+294602 & 109   & 242  & 0.52 & 0.23 & 6500 & 3150 & 73.5  & 0.016 & 0.32\\
CSS\_J081158.6+311959 & 101   & 206  & 0.47 & 0.23 & 6300 & 3400 & 61.3  & 0.018 & 0.33\\
\hline
\tablenotetext{a}{Main sequence binary model solution for the system.}
\tablenotetext{b}{WDMS binary model solution for the system.}
\end{tabular}
\end{center}
As noted in the text, these fits are based on velocities and fluxes from a single component in each system. 
They are thus only approximate. The fit temperatures for the secondary components are very uncertain and 
have been rounded to the nearest 100K.
Col. (1) Catalina ID.
Cols. (2) \& (3) Maximum radial velocities.
Cols. (4) \& (5) Masses of components.
Cols. (6) Fit effective temperatures of component.
Cols. (7) Input effective temperature based on observed spectral type.
Col.  (8) Orbital inclination.
Cols. (9) \& (10) Average radii.
\end{table*}

\begin{figure}[ht]{
\epsscale{1.0}
\plotone{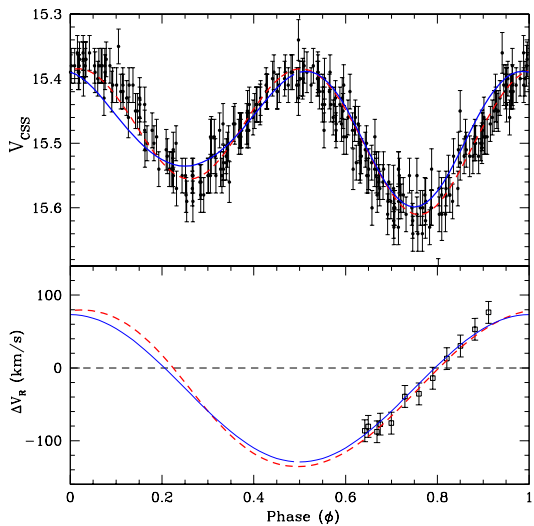}
\caption{\label{P200Vel}
The lightcurve and radial velocities variations of CSS\_J001242.4+130809.
The red dashed lines gives the dM+dM model fit to photometric and velocity
variation. The blue lines shows the best WD+dM fit. The fits are 
uncertain since only a single set of spectral features is observed.
}
}
\end{figure}

\begin{figure*}[ht]{
\epsscale{1.0}
\plottwo{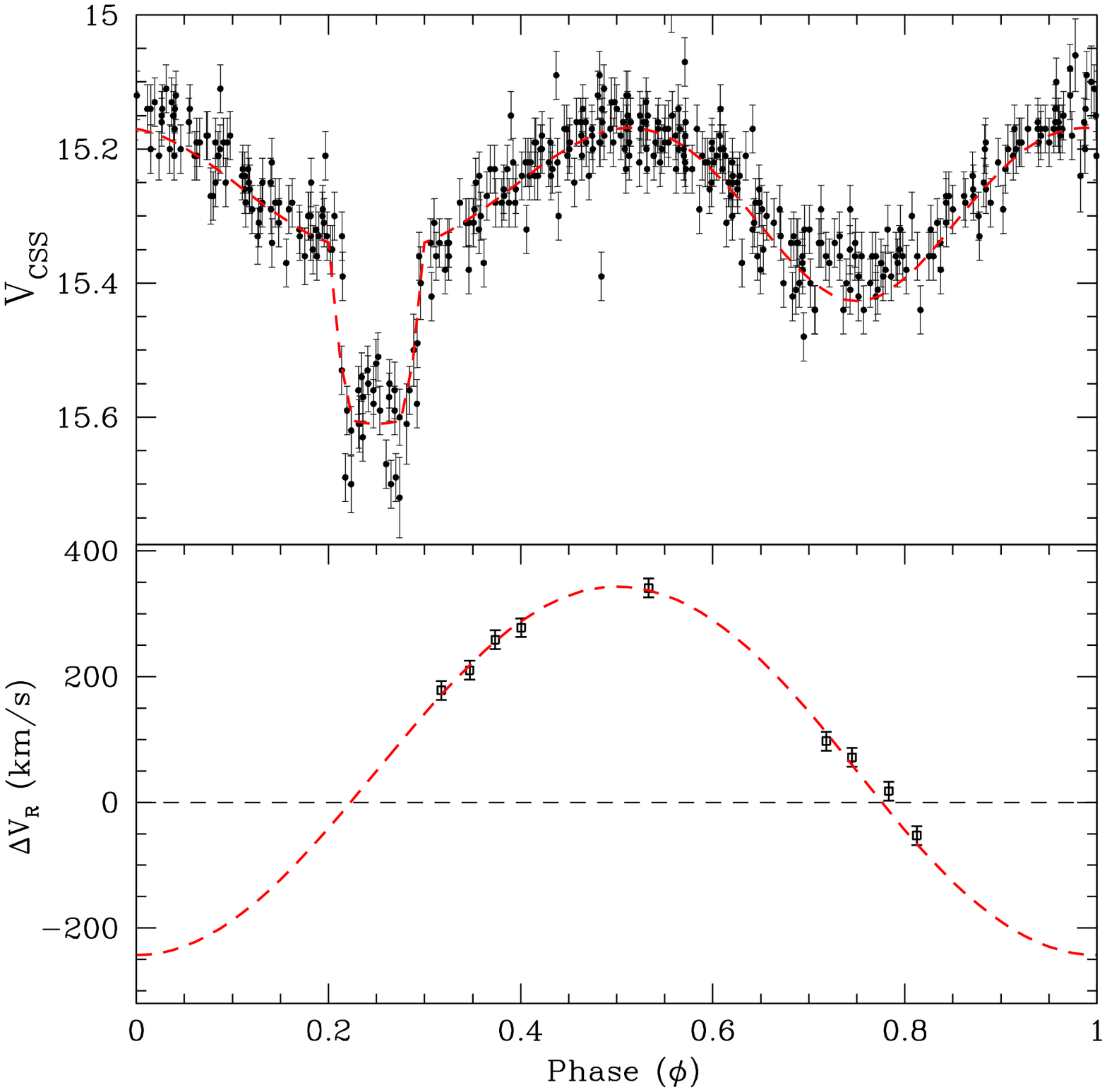}{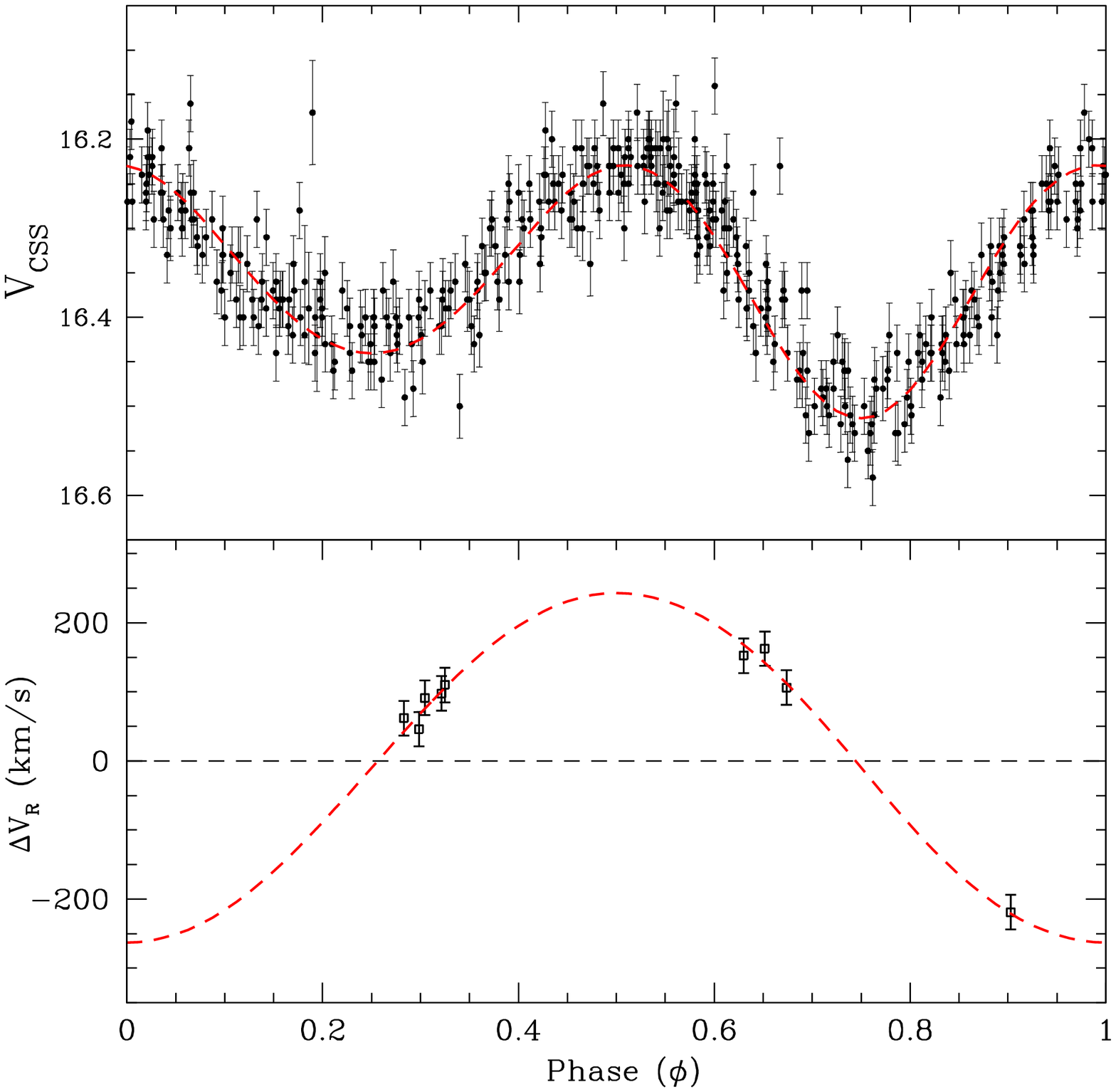}
\caption{\label{P200Vel2}
The lightcurves and radial velocities variations of ultra-short period
binaries. The left panel presents the results for CSS\_J090826.3+123648.
while the right panel shows the values for CSS\_J090119.2+114254.
The red dashed lines gives the model fit to photometric and velocity
variation. The fits are uncertain since only a single set of 
spectral features is observed.
}
}
\end{figure*}

\begin{figure*}[ht]{
\epsscale{1.0}
\plottwo{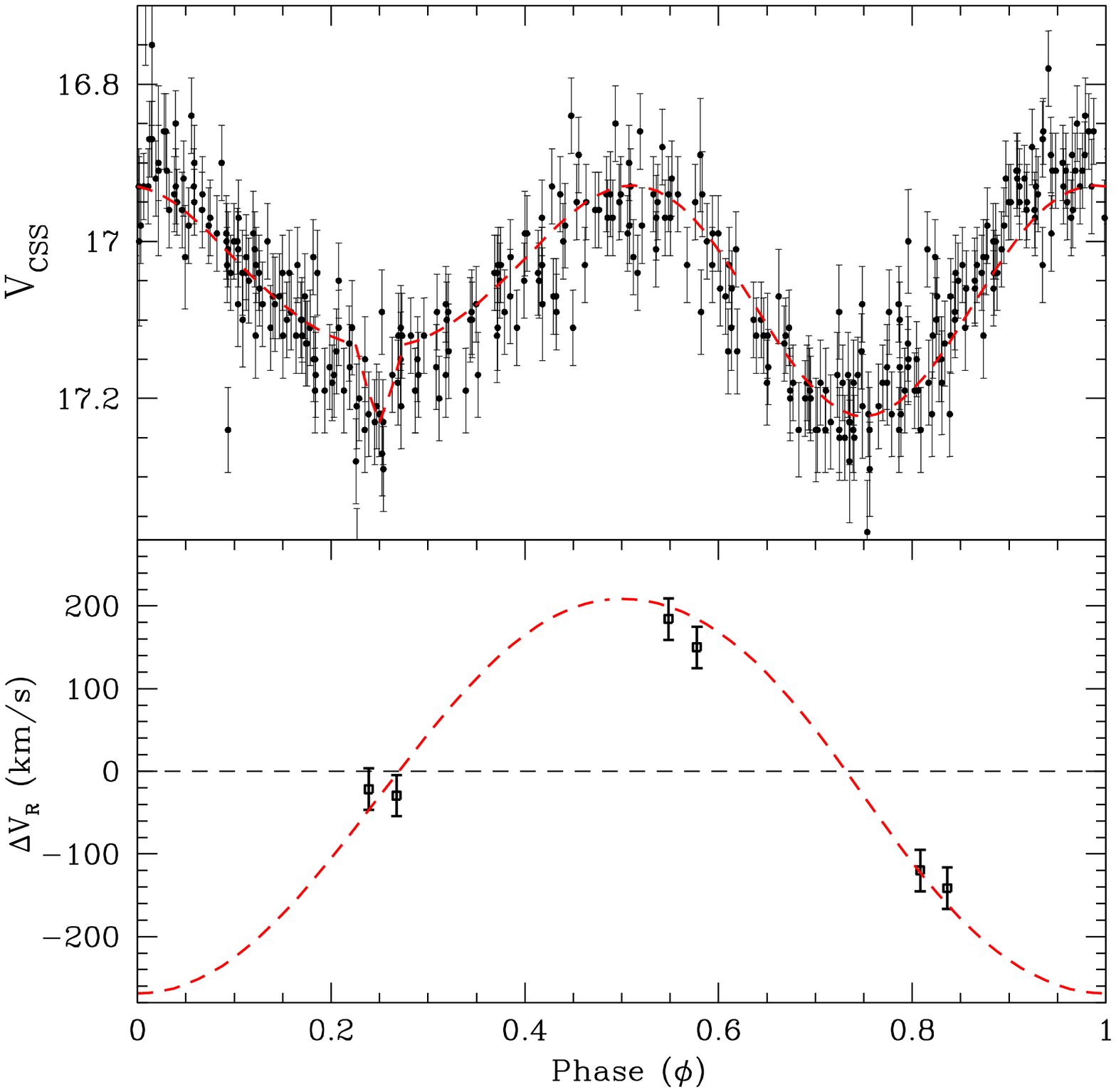}{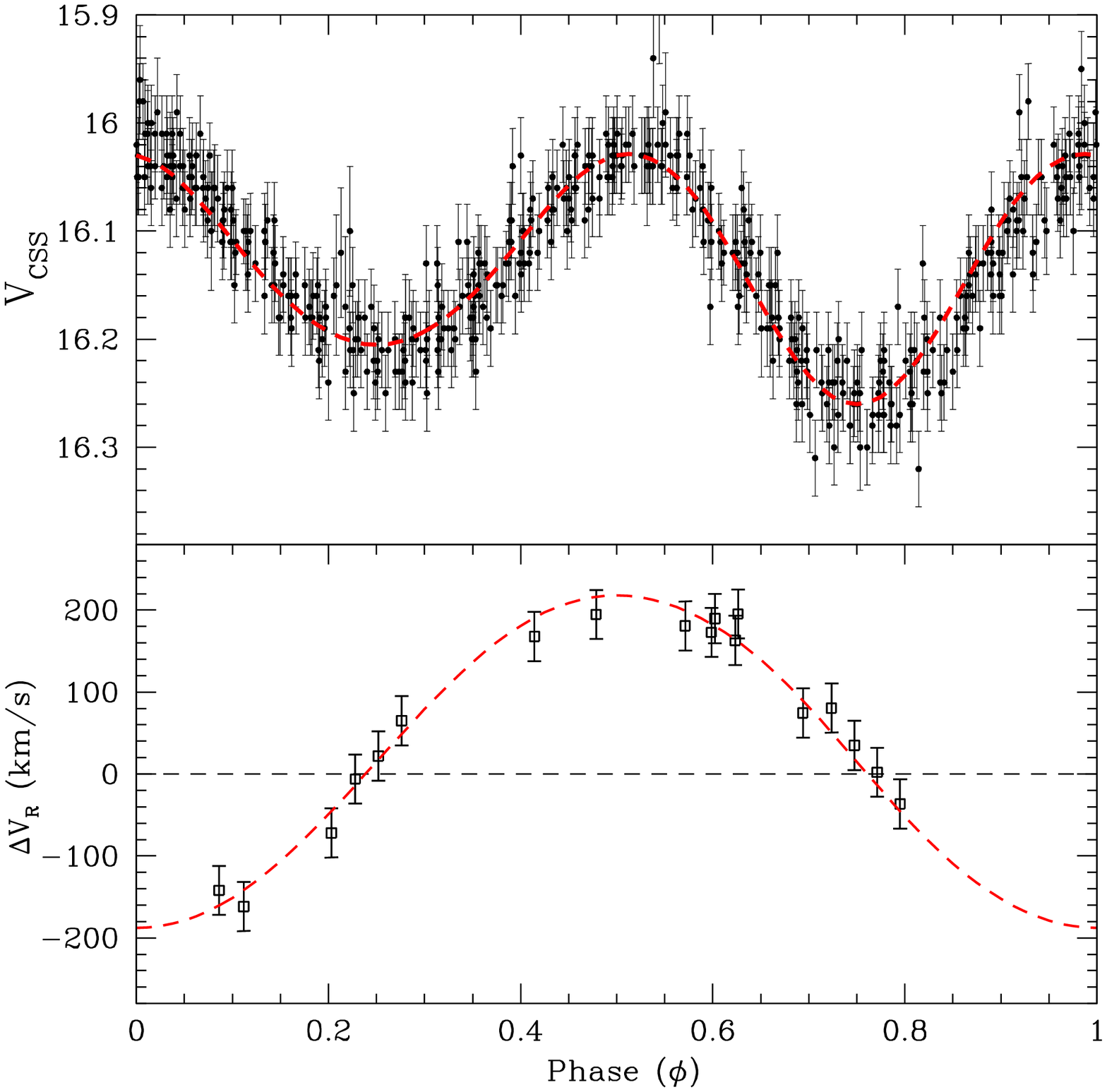}
\caption{\label{P200Vel3}
The lightcurves and radial velocities variations of ultra-short period
binaries. The left panel presents the results for CSSJ\_111647.8+294602.
while the right panel shows the values for CSS\_J081158.58+311959.
The red dashed lines gives the model fit to photometric and velocity
variation. The fits are uncertain since only a single set of 
spectral features is observed.
}
}
\end{figure*}

To determine the parameters for these binary systems we simultaneously fit the observed radial velocities and
lightcurves using the 
Nightfall\footnote{For details on Nightfall, see \url{http://www.hs.uni-hamburg.de/DE/Ins/Per/Wichmann/Nightfall.html}} 
software.  In Table \ref{TabFit}, we present the fit
parameters for the five systems and in Figures \ref{P200Vel}, \ref{P200Vel2} \& \ref{P200Vel3}, we present the light
curves and radial velocity variations. 
For each system we used the measured spectral type to estimate the approximate effective temperature based on 
Baraffe \& Chabrier (1996). The resulting binary system fits were sanity checked against radii and masses 
expected based on the updated models of M-dwarf presented Baraffe et al.~(1998). 
Nevertheless, as only one component is observed in each binary the system parameters in these fits are not very 
strongly constrained with the existing data.  Observations of both sources would put much stronger 
constraints on the components. In particular, the temperatures of the WDs are very poorly constrained 
due to their lack of flux from these sources.

From the plots it is clear that four out of five systems have velocity variations $> 100km/s$.  These four systems are
most consistent with WD+dM binaries containing a cool WD with a low-mass M-dwarf companion.  In fact, even without
detection of lines from both sources, the radial velocities of the emission lines are only consistent with origin 
from a low-mass component. We can be certain that the lines we observed are not from objects in the usual WD mass 
range ($0.4-0.8\rm M_{\sun}$).

For the fifth system, CSS\_J001242.4+130809, we found two possible solutions to the binary configuration.  This object
exhibits relatively small radial velocities, suggesting that a massive companion is not necessary to produce the
observed variations.  The first solution for this system consists of a pair of over-contact M-dwarfs orientated at a
high inclination to our line-of-sight ($40.5 \arcdeg$). The second solution suggests that the photometric and radial
velocity variations are due to a Roche lobe-filling M-dwarf at a much smaller inclination ($70.9\arcsec$). In the first
model, the radial velocity variations seen in balmer and Ca II emission lines would have to be due to the WD (unlike the
other systems), while in the second model, a distorted M-dwarf would give rise to the variations.  The difference
between the two fits can be seen from Figure \ref{P200Vel}.  The goodness-of-fit value for the dM+dM model was
$\chi^{2}_{r} = 1.1$, while for the WD+dM model it was $\chi^{2}_{r} = 1.8$. The dM+dM model is thus favored by this.
Additionally, the presence of emission from the M-dwarf seems more likely as it occurs for the other four systems.  If
the observed emission does originate from the WD, this would suggest that material from the M-dwarf is being accreted
onto the WD. For two similar WD+dM systems van den Besselaar et al.~(2007) and Bruch (1999) found that the emission was
due to the activity of the M-dwarf, rather than the WD.  Alternately, there are a few known examples of very low-mass
WDs in binaries (Brown et al.~2012).  Future high signal-to-noise observations, with increased phase coverage and
multi-band photometry, should be able to break the degeneracy between the two models.

\section{Discussion and Conclusions}

In Drake et al.~(2014) we classified 231 of the sources as short period eclipsing contact binaries and 136 as candidate
ellipsoidal variables.  Thus, among the $\sim 31,000$ contact and ellipsoidal binary candidates there are $\sim 370$ systems with
periods below the 0.22 day cutoff. Accounting for recent discoveries this work still increases the number of
ultra-short period eclipsing binaries and short period ellipsoidal systems by an order of magnitude.
However,our analysis demonstrates the difficulty in separating ultra-short period binaries containing pairs of
late main sequence stars, from those with cool WDs in WDMS systems.  For example, in cases where the observed flux from
a WDMS system is dominated the main sequence star, the presence of a WD may not be seen in optical spectra or
multi-colour photometry (such as provided by SDSS and GALEX). This suggests that such systems are best found using light
curves and confirmed using radial velocity variations.  However, in cases where the amplitude of variation is $>0.3$
mag, systems are likely to contain main sequence binaries, since the variation amplitude of a WDMS is limited by the
degree of distortion that the MS star can undergo before transferring material onto the WD (Parsons et al.~2013).

The PTF survey discovered three WD+dM systems with cool WDs in a search for planets transiting M-dwarfs (Law et al.
2012).  This survey covered $< 500$ sq deg for systems reaching $m_R \sim 18$. The three systems found have periods
between 0.35 to 0.45 days.  Based on this result the authors suggest that binaries with cool WDs are preferentially
found at large orbital radii, in contrast to the systems with hotter WDs presented by Rebassa-Mansergas et al.~(2012).
However, since the search undertaken by Law et al.~(2012) was designed to find planets transiting M-dwarfs they used the
standard box-fitting BLS technique of Kovacs et al.~(2002) to find variables.  The BLS technique is specifically
designed to discover small dips in otherwise featureless lightcurves, rather than other general types of periodic
behaviour. In particular, since short-period WD+dM systems exhibit large sinusoidal flux variations due to the distorted 
M-dwarf their lightcurves do not resemble transiting planets (Drake et al.~2003). Additionally, Parsons et al. (2013) 
have shown, only a small fraction of the hundreds of known WD+dM binaries exhibit the discrete eclipses that 
Law et al.~(2012) was sensitive to. Considering this fact and the relatively small area covered by the Law et al.~(2012) 
analysis compared to our survey, it is not surprising they did not find cool WD+dM systems with short periods. 
Indeed, in contrast to the Law et al.~(2012) results, Pyrzas et al.~(2012) discovered a cool (6000K) WD+dM system 
with a period of just 0.12 days. Nevertheless, as our analysis is limited to systems with periods $< 0.22$ days, 
we cannot constrain the possibility that there is indeed a much larger fraction of cool WD binaries at long periods.

The separation of WD+dM binaries from dM+dM systems is more difficult when the M-dwarfs are in the spectral range from 
M0V to M2.5V.  Such M-dwarfs have masses in the range 0.6-0.3$\rm M_{\sun}$ (Baraffe \& Chabrier 1996), and
absolute magnitudes $8 < M_V < 11$ (Kropa \& Tout 1997).  Since their masses overlap with the white dwarf mass distribution
(Kleinman et al.~2013), they give rise to the same radial velocity amplitudes as WDMS systems. 
Additionally, in WDMS binary systems, one does not expect to observe split narrow emission
lines that may be present in a main sequence binary pairs. Although, pairs of lines may not be present 
in many such systems.
Furthermore, as WDs have absolute magnitudes in the range $12 < M_V < 16$ (Bergeron et al. 1995, Andreuzzi et
al.~2002), cool white dwarfs can be many magnitudes fainter than their M-dwarfs companions and thus be undetectable.  
In order to distinguish these systems one has to investigate slight variations between over-contact and WDMS binary
solutions. However, if discrete eclipses are present this demonstrates that these systems are detached
rather than in contact.  

In this analysis we have found many of the short period sources with the ellipsoidal variable-type light curves.  As
noted above, while contact and detached binaries have different light curve shape, WD+dM binaries and over-contact
binaries with amplitudes $< 0.3$ mag often have very similar shapes.  Some of the systems exhibit excess $g - r$ flux
compared to main sequence binaries.  A few systems do exhibit $FUV-NUV$ colours (similar to those of previously known
CVs and WD+dM binaries).  However, these binary systems generally do not exhibit different $NUV-V_{CSS}$ colours compared
to main sequence eclipsing binaries or other periodic variables. Such systems therefore cannot be found through colour
selections with SDSS or GALEX, nor with individual SDSS spectra as analysed by Rebassa-Mansergas et al.~(2012).

The radial velocities for four of the five systems observed are sufficient to establish that the primaries are much more
massive than the M-dwarfs seen in the spectra. Instead, they are consistent with cool WDs.  For the remaining system, two
situations are possible; a WD+dM binary, or an M-dwarf pair. An M dwarf pair is favored due to the improved model fit,
relatively small radial velocities, and presence of narrow emission lines from the primary.

In comparison to our results, Nefs et al.~(2012) present a number of short-period systems that exhibit small amplitudes
and sinusoidal light curves.  Of the four Nefs et al.~(2012) sub-0.2 day systems, the lightcurve morphology of
07g-3-05744 best matches that of the ellipsoidal type variables we select.  Additionally, the extinction corrected SDSS DR9
colours for this source ($u-g=0.89$, $g-r=0.67$, $r-i=1.2$, $i-z=0.74$) strongly suggest that the object has a colour
excess. This system appears to be more likely an WD+dM binary than a main sequence pair.  No spectroscopic observations
were presented for this systems by Nefs et al.~(2012).  Two of the remaining three short period objects from Nefs et
al.~(2012) have SDSS u-band uncertainties greater than a magnitude, while the third is not detected and none of these
objects have measured radial velocities.  This makes it difficult to completely discount the possibility that the systems
contain cool white dwarfs. However, this possibility seems very unlikely for two of the systems, since they exhibit
regular detached binary lightcurves. In particular, 19h-3-14992 clearly exhibits two eclipses of similar span and 
differing depth.  Nefs et al.~(2012) confirms that this detached system contains an M-dwarf based on a WHT ISIS spectrum.
Although, the spectrum itself does not cover the $\lambda < 5000 \rm \AA$ region where a white dwarf might be seen, 
and there is no evidence for a companion in the spectrum presented.  

Comparing the number of short-period binaries discovered by Nefs et al.~(2012) one finds a much greater fraction
than found in our analysis. For examples, Nefs et al.~(2012) found 14 source with periods $< 0.22$ days from a sample 
of 262,000 sources, while we find 367 among 198 million sources. This difference can be partially explained by the 
much greater red sensitivity of the Nefs et al.~(2012) J-band data. In contrast, Norton et al.~(2011) found
only 15 similar short-period binaries sources among 30 million lightcurves using very well sampled, but shallower, 
SuperWASP data. Another reason for the relatively large number of Nefs et al.~(2012) sources is that the survey 
fields were taken much closer to the galactic plane. The CSS project avoided the plane because of crowding, yet 
as shown by Drake et al.~(2014) the contact binaries are strongly concentrated near the Galactic plane (as expected).

Using GTC, SDSS and Palomar spectroscopy we have confirmed the presence of M-dwarf systems among ultra-short period
eclipsing contact binaries.  The only other spectroscopically confirmed ultra-short period M-dwarf contact binary was
discovered by Davenport et al.~(2013).  This system, originally identified by Becker et al.~(2011), has a period of
$\sim 0.2$ days and clearly exhibits a sinusoidal light curve with an amplitude of $0.2$ mags. As with our ellipsoidal
variable selection, this object exhibits a slight deviation from a single sinusoid. 
The object's lightcurve shape is most consistent with our confirmed WD+dM systems. Furthermore, as with the WD+dM
binaries, the Davenport et al.~(2013) spectra lack evidence for a blue component yet exhibits an single $\rm H\alpha$
emission line. Davenport et al.~(2013) find that the masses of the components to be $M_1=0.54 M_{\sun}$ and $M_2=0.25
M_{\sun}$.  As noted above, such masses are consistent with a either an early M-dwarf or WD primary.  The
strongest evidence against this system being an WD+dM system, is the presence of faint pairs of Ca-I absorption lines 
at $\rm 6102\AA$ and $\rm 6122\AA$.  However, as Pyrzas et al.~(2012) have shown, WD+dM binaries also include metal
lines in their spectra.  In such cases the source of the lines has been attributed to WD accretion of M-dwarf wind. 
Since Davenport et al.~(2013) did not attempt to fit a WD+dM model, the exact nature of this systems still appears 
uncertain.

Our analysis firmly shows the existence of a population of contact binaries with periods $< 0.2$ days.  This result
suggests that current binary evolution models discounting the existence of these systems provide an incomplete picture
of the binary population.  Nevertheless, since M-dwarfs are very common and short-period systems very rare, it
is entirely possible that such systems only occur under special conditions. One possibility suggested by Nefs et al.~(2012)
is that such systems occur due to interactions in stellar triple systems.  Such interactions may not be uncommon since,
as noted by Rucinski, Pribulla \& van Kerkwijk~(2007), hierarchical triples are very common among short period 
binaries. 

The recent theoretical results of Jiang et al.~(2012) suggest that contact binaries are not found with masses less 0.61
$M_{\sun}$ and periods $< 0.2$ days due to their very short evolutionary times ($<$1 Gyr) that are caused by unstable
mass transfer. It is not possible to resolve the timescale of the unstable transfer presented by Jiang et al.~(2012), 
though the presented results suggest this is $<< 0.01 Gyr$.  It is possible that the small number of short-period 
contact binaries we detect might be those undergoing such transfer.  However, there is no observational evidence 
for this within the lightcurves.

In stark contrast to Jiang et al.~(2012), the Stepien (2006) binary models explain the lack of such systems
as being due to such systems not reaching contact within a Hubble time. Jiang et al.~(2012) explains the difference
between their model and those of Stepien (2006), as being due to a different assumption for the angular momentum loss 
rate. Given the apparently highly discrepant theories, the cause for the contact binary period limit remains
very poorly understood. Nevertheless, our discovery of contact systems below the 0.2 day limit should serve as 
an additional constraint for future binary models.

In the near future the Gaia mission (Perryman et al.~2001) and the VVV survey (Minniti et al.~2010) will begin to
harvest Galactic disk fields and are expected to find millions of periodic variables (Eyer et al.~2012; Catelan et
al.~2013). Although Gaia is only expected to reach stars to the same depth as CSDR1, with far fewer epochs, it is
expected to have ultra-precise photometry. This will greatly increase the accuracy with which over-contact and WD+dM
systems can be separated.  Likewise, the LSST survey will reach far greater depths than any existing wide-field survey
(Ivezic et al.~2008). The LSST will thus be able to probe M-dwarf binaries within a far greater volume than other
surveys, and thereby enable a better constrain the true frequency and period distribution of such systems.

\acknowledgements

CRTS and CSDR1 are supported by the U.S.~National Science Foundation under grant AST-1313422.
The CSS survey is funded by the National Aeronautics and Space Administration under Grant No.~NNG05GF22G issued through the
Science Mission Directorate Near-Earth Objects Observations Program. J.L.P. acknowledges support from NASA through
Hubble Fellowship Grant HF-51261.01-A awarded by the STScI, which is operated by AURA, Inc.  for NASA, under contract
NAS 5-26555.  Based on observations made with the Gran Telescopio CANARIAS (GTC), installed in the
Spanish Observatorio del Roque de los Muchachos of the Instituto de Astrofísica de Canarias, in the island of La Palma.
Support for M.C. and G.T. is provided by the Ministry for the Economy, Development, and Tourism's Programa Inicativa
Cient\'{i}fica Milenio through grant IC120009, awarded to Millennium Institute of Astrophysics (MAS), Santiago, Chile; 
by Proyecto Basal PFB-06/2007; and by Proyecto FONDECYT Regular \#1110326 and \#1141141.
SDSS-III is managed by the Astrophysical Research Consortium for the Participating Institutions of the SDSS-III
Collaboration Funding for SDSS-III has been provided by the Alfred P. Sloan Foundation, the Participating Institutions,
the National Science Foundation, and the U.S. Department of Energy Office of Science. The SDSS-III web site is
http://www.sdss3.org/.



\end{document}